\documentclass{ametsocV6.1}

\usepackage{soul}
\usepackage{gensymb}

\title{Global extreme heat forecasting using neural weather models}

\authors{Ignacio Lopez-Gomez,\aff{a,b}\correspondingauthor{Ignacio Lopez-Gomez, ilopezgp@google.com}
Amy McGovern,\aff{a,c}
Shreya Agrawal,\aff{a}
Jason Hickey,\aff{a}}

\affiliation{\aff{a}{Google Research, Mountain View, California}\\
\aff{b}{California Institute of Technology, Pasadena, California}\\
\aff{c}{University of Oklahoma, Norman, Oklahoma}
}

\abstract{Heat waves are projected to increase in frequency and severity with global warming. Improved warning systems would help reduce the associated loss of lives, wildfires, power disruptions, and reduction in crop yields. In this work, we explore the potential for deep learning systems trained on historical data to forecast extreme heat on short, medium and subseasonal timescales. To this purpose, we train a set of neural weather models (NWMs) with convolutional architectures to forecast surface temperature anomalies globally, 1 to 28 days ahead, at $\sim200~\mathrm{km}$ resolution and on the cubed sphere. The NWMs are trained using the ERA5 reanalysis product and a set of candidate loss functions, including the mean squared error and exponential losses targeting extremes.
We find that training models to minimize custom losses tailored to emphasize extremes leads to significant skill improvements in the heat wave prediction task, compared to NWMs trained on the mean squared error loss. This improvement is accomplished with almost no skill reduction in the general temperature prediction task, and it can be efficiently realized through transfer learning, by re-training NWMs with the custom losses for a few epochs. In addition, we find that the use of a symmetric exponential loss reduces the smoothing of NWM forecasts with lead time.
Our best NWM is able to outperform persistence in a regressive sense for all lead times and temperature anomaly thresholds considered, and shows positive regressive skill compared to the ECMWF subseasonal-to-seasonal control forecast after two weeks.}

\begin{document}
\maketitle

%
%
%

\statement

Heat waves are projected to become stronger and more frequent as a result of global warming. Accurate forecasting of these events would enable the implementation of effective mitigation strategies. Here we analyze the forecast accuracy of artificial intelligence systems trained on historical surface temperature data to predict extreme heat events globally, 1 to 28 days ahead. We find that artificial intelligence systems trained to focus on extreme temperatures are significantly more accurate at predicting heat waves than systems trained to minimize errors in surface temperatures, and remain equally skillful at predicting moderate temperatures. Furthermore, the extreme-focused systems compete with state-of-the-art physics-based forecast systems in the subseasonal range, while incurring a much lower computational cost.


%
%
%

%



\section{Introduction}
An important consequence of anthropogenic radiative forcing is the robust increase in heat wave days per year, both globally and at a regional level \citep{perkins_2020}. Heat waves pose a significant health risk, as evidenced by the more than 70,000 excess deaths that occurred during the 2003 European heat wave \citep{robine_2008}. More frequent heat waves will also lead to higher wildfire risk \citep{parente_2018, Ruffault2020}, stress on the power grid \citep{ke_2016}, and loss of agricultural crops \citep{Bras_2021}. These trends underscore the importance of developing effective mitigation strategies to reduce the negative impacts of extreme heat. A stepping stone in the development of such strategies are accurate forecasts with sufficient lead time \citep{Lin2022}. Current physics-based models, however, can only provide accurate forecasts of extreme heat events a few days in advance, which may not be sufficient to deploy effective mitigation strategies \citep{White2017, Wulff2019}.

There is mounting evidence of heat wave predictors on weekly to subseasonal timescales. These include large-scale quasi-stationary atmospheric Rossby waves \citep{Teng2013, Mann2018, white2021}, negative soil moisture anomalies \citep{Vautard2007, Benson2021}, and anomalous Pacific sea surface temperature (SST) gradients in the case of North American heat waves \citep{Deng2018, Miller2021}. Recently, \citet{Miller2021} used a linear regression model based on empirical orthogonal functions of the North Pacific SST and soil moisture over the United States to predict the weekly frequency of extremely warm days in the United States, 1 to 4 weeks ahead. They show that their statistical model outperforms the operational NCEP CFSv2 model in this task over the eastern United States after the second week, which suggests that purely data-driven forecasting may provide a path forward in extreme heat prediction beyond the 10-day horizon.

In this context, deep neural networks represent a natural extension of the data-driven approach, given their remarkable success in image segmentation and forecasting tasks \citep{unet, sonderby2020metnet, Ravuri2021}. Different methods to classify heatwaves leveraging deep learning have recently been proposed.  \citet{Chattopadhyay2020} trained a capsule neural network on midtropospheric geopotential and surface temperature from a large ensemble of climate model runs to classify several future days into 5 different classes, each representing either a specific heat wave pattern over North America, or the absence of extreme temperatures. \citet{jacques_dumas_2021} trained a convolutional neural network on wavenumber space to forecast the occurrence of heat waves in France with a 15-day lead time. They used the same predictors as \citet{Chattopadhyay2020}, and data from a $1000$-year cyclic climate model simulation.

These studies showcase the potential of deep learning to forecast extreme heat events as a classification problem, in particular regions, and trained on very large datasets sampled from quasi-stationary distributions. Here, we tackle some of the practical questions left unanswered by previous work:
\begin{itemize}
    \item Can deep learning models forecast extreme heat events when trained on limited historical data? The use of observations or reanalysis data is crucial for systems to improve upon existing physics-based models, since deep learning models trained solely on synthetic data will at best inherit the biases of the numerical models they are trying to substitute.
    \item Can general purpose neural weather models (NWMs) be used to predict extreme heat events? By general purpose, we refer to deep learning systems trained to minimize errors in the underlying fields, such as temperature, and not on extreme classification explicitly \citep[e.g.,][]{Rasp2020, Weyn2020, Pathak2022, Keisler2022}.
    \item Can NWMs improve their extreme prediction skill through the use of custom losses, while retaining their skill in a general weather forecasting setting? 
\end{itemize}

To answer these questions, we frame extreme heat prediction as a regression problem and restrict our training data to a large subset of the ERA5 reanalysis product.  Framing the forecast problem as a regression task bridges the gap with NWMs that act as time integrators, and require reliable previous forecasts as inputs to generate a new prediction \citep{Weyn2020}. The regression problem is also more robust to the definition of heat waves, diverse in the literature \citep[e.g.,][]{Chattopadhyay2020, Wulff2019, Miller2021}, and allows learning about the nuances of target states that may otherwise be masked under the same class in a classification problem. The advent of skillful regression-based NWMs would democratize the use of ensemble-based weather forecasting for targeted applications, which requires enormous computational resources when realized through state-of-the-art physics-based models \citep{Palmer2017}. In contrast, deep learning systems designed to forecast a few fields of interest (e.g., surface temperature) only incur high computational costs during training, but not during inference \citep{Scher2021, Weyn2021}.

To address the second question, we make use of a state-of-the-art convolutional architecture on the cubed sphere, following \citet{Weyn2020}, so that results can be extrapolated to similar NWMs described in the literature. Finally, to explore the last question, we compare forecasts of NWMs trained with the general purpose mean squared error loss to forecasts from NWMs trained to minimize custom losses that emphasize extremes. All results presented are contextualized through comparison with the ECMWF S2S operational forecast system \citep{Vitart2017}.

The paper is organized as follows. In Section \ref{sec:2_problem_statement}, we define the forecasting task and describe the data and losses used to train the NWMs. In Section \ref{sec:3_model_arch}, the model architecture is discussed. Section \ref{sec:4_results} explores the skill of NWMs trained using different loss functions in tasks varying from extreme heat to general surface temperature prediction, including example forecasts for the 2017 Iberian heatwave and the 2021 western North American heatwave. In Section \ref{sec:interpretability}, the relevance of the different NWM inputs is explored using integrated gradients \citep{Sundararajan2017}. Finally, Section \ref{sec:6_discussion} ends with a discussion of the results and potential future research directions.

\section{The forecasting problem}
\label{sec:2_problem_statement}
The forecasting task, given a set of observed input and target pairs $(x, y)$ with $y \in \Omega \subset \mathbb{R}^d$ and $x \in V \subset \mathbb{R}^q$, can be framed as the following minimization problem \citep{LopezGomez2022},
\begin{equation}\label{eq:min_problem}
    \theta^* = \arg\min_{\theta} \int_\Omega L(\Psi(\theta, x), y)dy,
\end{equation}
where $\Psi: \mathbb{R}^p \times \mathbb{R}^q \rightarrow \mathbb{R}^d$ is a mapping from parameter and input space to target space, $\theta \in \mathbb{R}^p$ are model parameters, $x \in \mathbb{R}^q$ is an input data vector, and $L(\cdot, \cdot)$ is some loss that we seek to minimize over a target set $\Omega \subset \mathbb{R}^d$. In this paper, we consider $\Psi$ to be a convolutional neural network operating on a gnomonic equiangular cubed sphere, described in detail in Section \ref{sec:3_model_arch}. Architecture exploration for NWMs is an active area of research \citep{Pathak2022, Keisler2022}, but it is not the emphasis of this work, so we keep the architecture $\Psi$ fixed. Instead, we are interested in comparing the usefulness of NWMs $\Psi(\theta^*, \cdot)$ that result from the minimization (\ref{eq:min_problem}) when varying the definition of the loss. Finally, the NWM parameter vector $\theta^*$ is obtained through mini-batch gradient descent, using the Adam algorithm \citep{kingma2017adam}.

\subsection{Data, predictors and targets}
\label{subsec:2a_data}
\begin{figure}[h]
 \begin{center}
 \noindent\includegraphics[width=19pc]{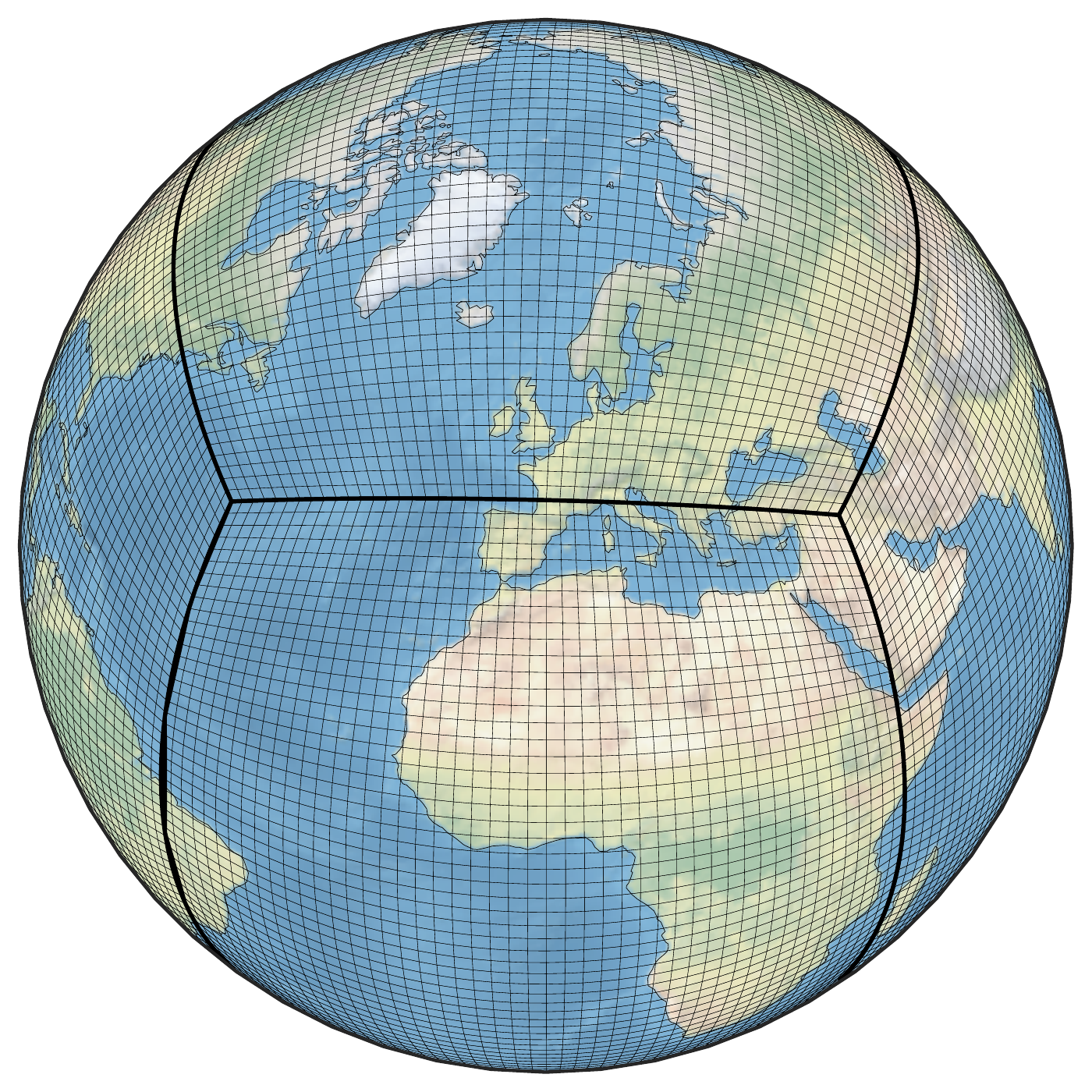}\\
 \caption{Depiction of the gnomonic cubed sphere grid onto which predictors and targets are projected. The cubed sphere is composed of 6 faces with $48^2=2304$ cells each.}
 \label{fig:cubed_sphere}
 \end{center}
\end{figure}

The targets $y$ are constructed from the daily average of the standardized climatological anomalies of the temperature 2$~\mathrm{m}$ above the surface ($T_{2m}$), which we denote $\tilde{T}_{2m}$. We recognize that air temperature is a suboptimal indicator of heat-related illness \citep{Xu2016, Heo2019}, but it enables comparison with other models in the literature \citep{Weyn2021, Wulff2019, Lin2022}. Each target vector $y$ includes $\tilde{T}_{2m}$ for a set of lead times $\tau=1, \dots, \tau_l~\mathrm{days}$ and for all tiles of a gnomonic equiangular grid of Earth's surface \citep{Ronchi1996}. Thus, the target size $d$ is the flattened length of the tensor $y_{cs} \in \mathbb{R}^{\tau_l\times f\times h \times w}$, where $f=6$ is the number of faces of the cubed sphere, and $h=48,~w=48$ are the number of meridional and zonal tiles of each face, respectively. The grid is shown in Figure \ref{fig:cubed_sphere} for reference; the surface area of each tile is approximately $192^2~\mathrm{km}^2$. To assess the skill of the model from the short to the subseasonal range, we aim to predict temperature anomalies for the next $\tau_{l}=28$ days.

The inputs $x \in V \subset \mathbb{R}^q$ contain $v$ daily-averaged surface fields on the cubed sphere, such that $q$ is the flattened length of the tensor $x_{cs} \in \mathbb{R}^{v\times f\times h \times w}$. Here $v=\tau_pv_t + v_i$ is the number of input fields, and $v_t,v_i$ are the number of time-dependent and independent fields, respectively. For time-dependent fields, daily averages from the last $\tau_p=7$ days are included.  We consider as time-dependent fields the geopotential height and potential vorticity at $300, 500$ and $700~\mathrm{hPa}$, the temperature $T_{2m}$, the top net outgoing long-wave radiative flux (OLR), and the volume of soil water. To these fields we append as auxiliary time-independent variables the latitude, longitude, topography and a land-sea mask. We also include the present insolation and date as inputs. The full list of predictors is summarized in Table \ref{table:predictors}.

The choice of predictors is informed by studies linking extreme heat events to mid-tropospheric geopotential height \citep{Teng2013, Mann2018}, soil moisture \citep{Vautard2007, Benson2021} and large-scale phenomena with characteristic OLR signatures, like the Madden-Julian Oscillation \citep[MJO,][]{jacques_coper2015, Maloney2019} or the Boreal summer intraseasonal oscillation \citep{Lin2022}. In addition, we include $T_{2m}$ to learn about transport processes such as advection; and its standardized anomaly to facilitate improving upon a persistence model. Both $T_{2m}$ and $\tilde{T}_{2m}$ capture the signature of large-scale oscillations such as El Niño (ENSO) or the North Atlantic Oscillation \citep{Ogi2003, Wang2011, Wright2014}.

\begin{table}[h]
\caption{Predictors used for the temperature anomaly forecasting task. Anomalies are standardized (std.) with respect to climatology. Heights are specified as above (a.g.) and below (b.g.) ground level.}\label{table:predictors}
\begin{center}
\begin{tabular}{ccccrrcrc}
\topline
Predictor & Levels\\
\midline
 Temperature anomaly (std.) & $2~\mathrm{m}$ (a.g.)  \\
 Temperature & $2~\mathrm{m}$ (a.g.)  \\
 Geopotential height anomaly (std.) & $300, 500, 700~\mathrm{hPa}$  \\
 Geopotential height & $300, 500, 700~\mathrm{hPa}$  \\
 Potential vorticity anomaly (std.) & $300, 500, 700~\mathrm{hPa}$  \\
 Potential vorticity & $300, 500, 700~\mathrm{hPa}$  \\
 Outgoing longwave radiation & Top of atmosphere \\
 Incoming shortwave radiation & Top of atmosphere \\
 Surface soil moisture & $0-7, 7-28~\mathrm{cm}$ (b.g.) \\
 Deep soil moisture & $28-100, 100-289~\mathrm{cm}$ (b.g.) \\
 Topography & N/A \\
 Land-sea mask & N/A \\
 Latitude, longitude & N/A \\
 Date & N/A \\
\botline
\end{tabular}
\end{center}
\end{table}

All data are daily averages from the ERA5 reanalysis product \citep{era5}, downloaded at $2^{\circ}\times2^{\circ}$ resolution from the Copernicus Climate Data Store (CDS) and projected onto the cubed sphere following \citet{ulrich_2015a, ulrich_2015b}. The climatology is computed for the time period $1979-2019$. All days of the year, not only the summer days, are used to train the model. We do this to learn about physical processes that are season independent, like advection. When framed as a classification task, extreme heat prediction may require undersampling of non-extreme samples during training \citep{jacques_dumas_2021}. Here, we make use of all available information with no explicit undersampling; class imbalance is dealt with through the use of custom losses as discussed in Section \ref{sec:2_problem_statement}\ref{subsec:2b_losses}. In addition, we perform a sequential split of the data into training (1979-2012), validation (2013-2016) and test sets (2017-2021). The limited amount of historical data available means that the influence of longer modes of climate variability (e.g., the Pacific Decadal Oscillation) is unlikely to be robustly captured. Furthermore, a shift in the target distribution from training to testing sets is implicit with this split, due to climate change \citep{white2021, chan2020}. This shift in the temperature distribution is, however, representative of situations in which a warning system might be used in practice, since both data-driven and NWP models are calibrated using historical observations.

\subsection{Losses considered}
\label{subsec:2b_losses}

Unless the optimal parameter vector $\theta^*$ is able to yield a perfect model (i.e., $\Psi(\theta^*, x)=y \;\; \forall (x, y)$), the optimum will depend on the definition of the loss. This is the case in the extreme heat prediction task, since the chaotic nature of the atmosphere precludes a perfect forecast of trajectories from inexact initial conditions \citep{ Lorenz1969b, LORENZ1969, Slingo2011}. For this reason, we can expect models that minimize generic losses to be suboptimal for the extreme prediction task.

To study the potential benefits of training NWMs on losses targeting extreme prediction, we consider two losses: the mean squared error (MSE) and a custom loss $L_e$ based on the exponential of targets and forecasts,
\begin{equation}\label{mse_exp}
    L_e(y', y) = a\mathrm{MSE}( e^{y'}, e^{y}) + b\mathrm{MSE}( e^{-y'}, e^{-y}),
\end{equation}
where $y$ are the targets and $y'$ are the forecasts. In equation \eqref{mse_exp}, the choice $(a,b)=(1,0)$ emphasizes the correct prediction of positive extremes, $(a,b)=(0,1)$ emphasizes negative extremes, and the midpoint $(a,b)=(0.5, 0.5)$ emphasizes both extremes. In the context of temperature anomaly prediction, these losses emphasize heat waves, cold spells and extreme deviations from climatology, respectively. This custom loss is motivated by the reported success of neural networks in extreme prediction tasks when using target transformations involving the softmax and softmin functions \citep{Qi_Majda_2020}. However, we exclusively take the numerator of the suggested transformations because the softmax and softmin functions are invariant under translational shifts \citep[Appendix A,][]{LopezGomez2020}, which in this application means that climatological biases would not be penalized. 

In the following, we denote models trained with the loss \eqref{mse_exp} as HeatNet for $(a,b)=(1,0)$ and ExtNet for $(a,b)=(0.5,0.5)$. The model trained with the MSE loss, representative of general neural weather prediction systems \citep{Rasp2020, Weyn2020, Weyn2021}, is denoted GenNet. We trained HeatNet, ExtNet and GenNet models using a hyperparameter search over the learning rate, and the magnitude of L1 and L2 norm regularization. All models were trained until they started overfitting to the training set, evidenced by an increase in validation loss persistent over many epochs. Notably, our best HeatNet and ExtNet models were obtained through transfer learning, by retraining our best GenNet model for a few ($<3$) epochs on the custom exponential loss. This implies that any performance improvements of HeatNet or ExtNet with respect to generic models trained on the MSE loss can be realized efficiently through transfer learning from the original models. Our transfer learning methodology relies on early stopping to retain an inductive bias toward the GenNet parameters \citep{yosinski_2014, li_2018}. The implementation of other transfer learning techniques, like Bayesian regularization toward the original (i.e., prior) model parameters \citep{li_2018, Inubushi2020}, may result in further skill improvements and will be explored in the future.

\section{Neural weather model architecture}
\label{sec:3_model_arch}
We employ a convolutional architecture to construct the neural network $\Psi$, which maps the input fields at all past times $\tau = -6, \dots, 0~\mathrm{days}$ to the daily-averaged temperature anomaly $\tilde{T}_{2m}$ at all lead times $\tau=1, \dots, 28~\mathrm{days}$. Consistent with our projection of the data, convolutions are performed on each of the cubed sphere faces, using halo exchange at the borders \citep{Weyn2020}. Kernel weights are shared among all 4 equatorial faces, and a different set of kernel weights is used for the polar faces. This enables learning about different processes governing on one hand tropical and sutropical dynamics, and on the other mid- and high-latitude dynamics. The northern polar face is mirrored before each convolution to align cyclonic and anticyclonic motions in each hemisphere, following \cite{Weyn2020}.

\subsection{Receptive field}
\begin{figure}[h]
 \noindent\includegraphics[width=39pc]{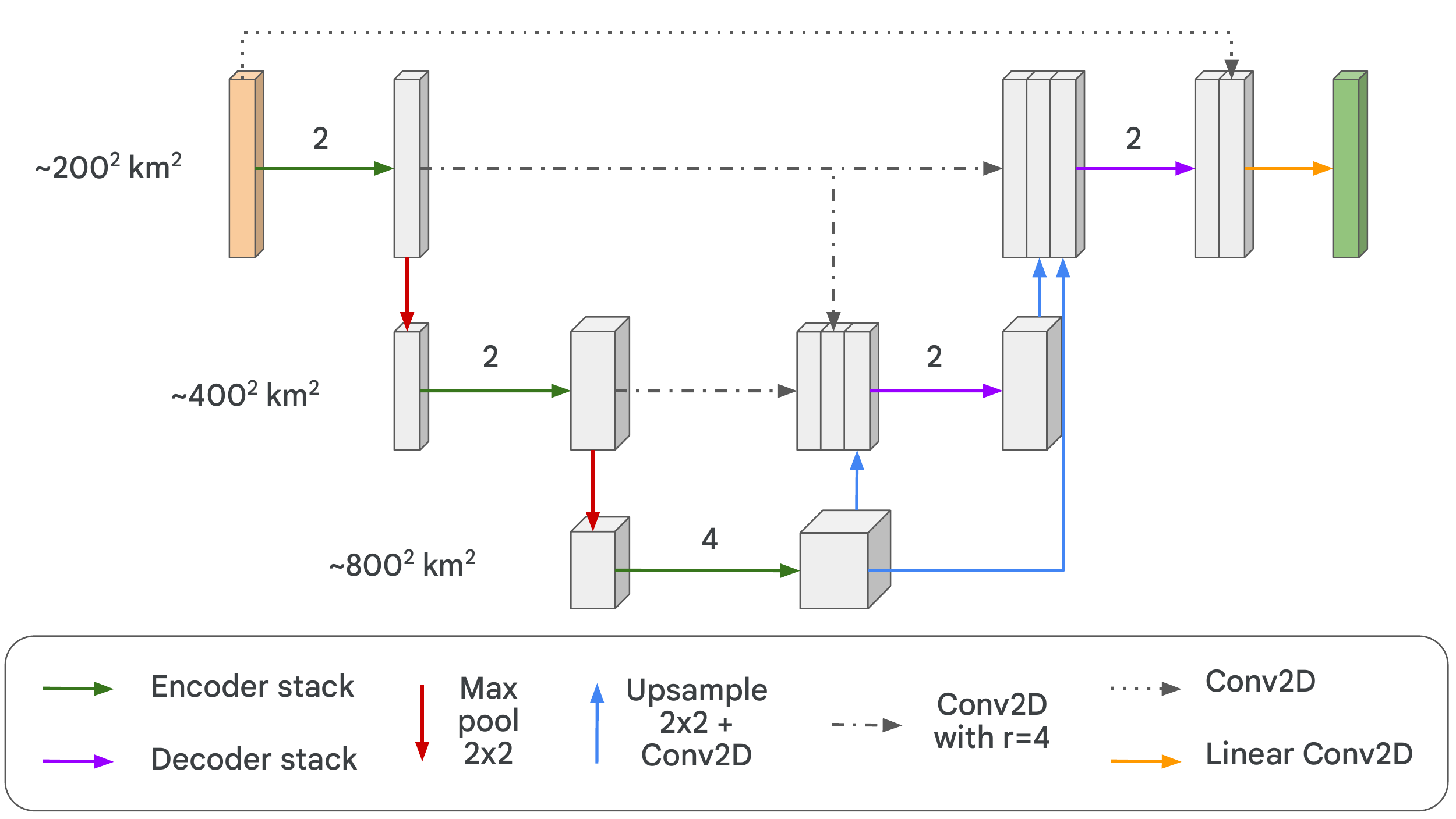}\\
 \caption{Neural weather model architecture, modified from a UNet 3+ architecture \citep{Huang2020}. The number of layers of each encoder and decoder stack are as indicated in the schematic. Encoder convolutional layers have geometrically-growing dilation factors $r=2^l$, where $l=0,1,\dots$ is the layer number within the stack from inputs to outputs, and decoder layers have dilation factors $r=8$ and $16$. The layers connecting same-level encoders and decoders have convolutions with 32 filters and dilation factor $r=4$. All other layers have dilation factor $r=1$, and all layers have convolutional kernels of size $3\times3$. Same-level layers implement 32, 64 and 128 filters in the first, second and third levels, respectively.}
 \label{fig1:model_arch}
\end{figure}

Due to the non-recurrent nature of the architecture and the lead times considered, it is crucial to achieve a fully receptive field if we want to capture long-range dependencies and teleconnections \citep{Espeholt2022}. A fully receptive field is realized through two design characteristics of the proposed architecture, which is sketched in Figure \ref{fig1:model_arch}. The first one is the use of dilated convolutions, which rapidly increase the receptive field of any location on the cubed sphere as information traverses the network \citep{dilated_conv}. The second one is the use of a UNet-type architecture \citep{unet} with 3 resolution levels going from the data resolution to the synoptic scale: $\sim200^2, \; \sim400^2$ and $\sim800^2~\mathrm{km}^2$. Coarser-resolution levels increase the receptive field proportionally to their downsampling rate, allowing to achieve larger receptive fields with fewer layers.

\subsection{Encoder and decoder architecture}
The architecture of our model is based on the UNet 3+ architecture \citep{Huang2020} with a few modifications. All nonlinearities consist of Parametric Rectified Linear Units  (PReLUs) that share parameters across all dimensions except the channels \citep{He2015}. We use dilated convolutions, as previously mentioned, with dilation factors $r$ that increase geometrically with network depth at every resolution level. The first two levels have encoder and decoder stacks with 2 layers each, while the synoptic-scale level is composed of an encoder stack with 4 layers.

Each encoder layer $l=0,1,\dots$ applies 2D $3\times3$ dilated convolutions with dilation factor $r=2^{l}$ and a PReLU nonlinearity. The decoder layers in the first two levels apply 2D $3\times3$ convolutions with dilation factors $r=2^3, \; 2^4$, and are each followed by a PReLU nonlinearity. In addition, we include a nonlinear skip connection at the finest resolution level to easily capture persistence. Downsampling between levels is performed using max-pooling. In the decoder, upsampling is followed by 2D $3\times3$ convolutions. The number of layers per level was obtained through cross-validation from a small set of architectures that achieved a full receptive field.

All layers at $200^2~\mathrm{km}^2$ are composed of 32 convolutional filters, and layers at $400^2~\mathrm{km}^2$ and $800^2~\mathrm{km}^2$ resolution apply 64 and 128 filters to their inputs, respectively. The skip connections between encoder and decoder stacks, as well as the upsampling layers, have 32 filters each. In each layer of the network, we use two independent convolutional kernels for each filter: one covering all 4 equatorial faces of the cubed sphere, and the other covering the polar faces. As shown in Section \ref{sec:4_results}, there is no discernible imprint of the cubed sphere edges on the model forecasts. In the end, the model architecture has about $1.8$ million parameters, halfway between the complexity of the models used in \citet{Weyn2020} and \citet{Weyn2021}.

\section{Results}
\label{sec:4_results}

\subsection{Reference models considered}
\label{sec:models_considered}

We assess the skill of HeatNet, ExtNet and GenNet against persistence and the European Centre for Medium-Range Weather Forecasts (ECMWF) subseasonal-to-seasonal (S2S) forecast system \citep{Vitart2017}. The ECMWF S2S system is an operational model that provides real-time 46-day forecasts twice a week. For dates in the test set ($2017-2021$), real-time forecasts used ECMWF's IFS cycles CY43R1, CY46R1 and CY47R2 \citep{p17118, p19309, p19749}. The S2S system employed 91 vertical levels until May 2021, and 137 levels after that. All versions are coupled to an ocean model at $0.25\degree$ resolution with an interactive sea-ice model, and use a triangular-cubic-octahedral horizontal discretization with $16~\mathrm{km}$ resolution for days $1-15$, and $32~\mathrm{km}$ after that \citep{ifs_grid}.

In order to compare against our deep learning systems, the ECMWF forecasts are bilinearly interpolated to the same resolution as the input ERA5 data ($2\degree$), subtracting the same ERA5 mean climatology to produce climatological anomalies. Then, the results are mapped to be cubed sphere using the conservative remapping of \citet{ulrich_2015a} and \citet{ulrich_2015b}; all skill metrics are computed on this grid. To assess potential errors due to the spherical harmonics truncation employed by ECMWF's Meteorological Archival and Retrieval System (MARS), we downloaded the forecasts at $2\degree$ and $0.25\degree$ resolutions, and compared the forecasts after bilinearly interpolating the latter to the $2\degree$ grid. The root mean squared difference between the forecasts is $\sim 10^{-3}~\mathrm{K}$, much lower than typical forecast errors.

For all comparisons in this study, we employ both the real-time daily averaged ECMWF control and perturbed ensemble forecasts \citep{Vitart2017}. Model drift is removed from the real-time ECMWF forecasts using 660 reforecasts covering the past 20 years and initialized from ERA5 data \citep{Vitart2017}.  Comparisons with the ECMWF control assess the skill of NWMs against a deterministic "best-guess" physics-based forecast. The ECMWF S2S ensemble prediction system employs 50 additional ensemble members, perturbing both their initial conditions and model physics to capture forecast uncertainty \citep{Buizza1999}. Operational warning systems typically use perturbed ensembles, which have been shown to yield a higher economic value than high-resolution deterministic forecasts \citep{Richardson2000, Palmer2017}. For this reason, we include the ECMWF ensemble mean forecast for comparison. Information beyond the first moment of the ensemble statistics is also valuable \citep{Molteni1996, Zhu2002, Palmer2017}. However, we limit our comparison to the ensemble mean in this study, since we only consider NWM point forecasts. Even though our models yield a single deterministic output, direct NWM forecasts more closely resemble an ensemble mean prediction than a physical trajectory of the system; this interpretation is supported by the results in Sections \ref{sec:4_results}\ref{subsec:4a_summer_land}-\ref{subsec:pnw_heatwave}.

Two additional points should be considered when interpreting the relative skill of the ECMWF forecasts. First, the real-time ECMWF system is initialized from the operational IFS analysis, not ERA5, which leads to reduced accuracy at short lead times. Second, the native resolution of the ECMWF system is higher than the resolution of the NWMs. This is both an asset and a liability when evaluating pointwise objective scores; higher resolution reduces structural model errors, but inevitable errors in the timing and location of sharper resolved features can result in lower skill \citep{Mass2002}. Nevertheless, negative impacts of resolution on forecast skill are reduced in our study through the smoothing induced by bilinear interpolation and daily averaging \citep{Accadia2003}.

\subsection{Forecast skill for summer over land}
\label{subsec:4a_summer_land}
Although we train the NWMs using global data from all seasons, we evaluate here the performance of the forecast systems exclusively for summer over land, where heat wave prediction is most relevant. We define summer as the June-July-August trimester for the Northern Hemisphere and December-January-February for the Southern Hemisphere.

To assess model skill during increasingly hot summer days, we evaluate forecasts using two different temperature anomaly percentiles: the 75th (hot) and 95th (extremely hot) percentiles. Setting these thresholds allows assessing the forecast systems as binary classifiers. When evaluating regressive skill, conditioning on the target distribution can confront forecasters with the dilemma of overforecasting a rare event to improve their scores. There is no obvious way to avoid this problem when evaluating the regressive skill of deterministic forecasts at predicting extremes \citep{Lerch2017}. We verified that this dilemma is not a concern for the models we evaluate, after their global bias is subtracted, since they either underpredict extreme anomalies (NWMs), or are well-calibrated (ECMWF control); results conditioned on the union of forecast and target values, which account for false alarms, are included in Appendix B.

The regressive skill of the models is characterized in this study through the debiased root mean squared error (RMSE$_d$) and the centered anomaly correlation coefficient (AnCC) of standardized temperature anomalies. The RMSE$_d$ is defined as the RMSE of forecasts with respect to targets after removing the global mean bias per lead time of forecasts with respect to targets in the entire test set. We choose to debias the forecasts to prevent forecast bias from positively affecting the skill metric, since the mean target above the temperature thresholds is nonzero \citep{Lerch2017}. The subtracted bias is shown for all models in Figure \ref{fig:physical_metrics_general} for reference; it is clear that subtracting the bias prevents HeatNet from hedging. 

The centered anomaly correlation coefficient for a given lead time $i$ is defined as \citep{Wulff2019}
\begin{equation}\label{eq:centered_acc}
    \mathrm{AnCC}_i = \dfrac{\sum_{k=1}^{N_i}(y_{ik}-\bar{y}_i)(y'_{ik}-\bar{y}_i')}{\sqrt{\sum_{k=1}^{N_i}(y_{ik}-\bar{y}_i)^2\sum_{k=1}^{N_i}(y'_{ik}-\bar{y}_i')^2}},
\end{equation}
where $\bar{y}_i\in \mathbb{R}$ is the temporal and spatial average of the target temperature anomaly $\tilde{T}_{2m}$ at lead time $i$ for summer over land over the entire test set; $y_{ik}\in \mathbb{R}$ are individual values of $\tilde{T}_{2m}$ at lead time $i$ and at a given location and summer day, indexed by $k$; and the sums are over the $N_i$ summer targets $y_{ik}$ above the considered standardized temperature anomaly threshold. The forecast counterparts $\bar{y}_i'$ and $y_{ik}'$ are defined similarly based on the forecast temperature anomaly, but the sum over indices $k$ is still conditioned on the anomaly threshold of the targets. The AnCC is a useful metric of the potential of a forecast system, measuring the correlation between the target and forecast outputs \citep[Chapter 9, ][]{Wilks2019}. Our definition (\ref{eq:centered_acc}) takes into account the dynamic temperature anomalies, filtering out the thermodynamic shift of temperature anomalies over land with respect to the 1979-2019 climatology due to global warming, as well as lead-time dependent model biases. This is not the case for the non-centered anomaly correlation coefficient, which does not filter out forecast biases with respect to climatology \citep[e.g.,][]{Weyn2020}.

The classification skill of the models is evaluated through the extremal dependence index \citep[EDI,][]{edi_ferro2011} and the equitable threat score (ETS). The EDI is defined as
\begin{equation}\label{eq:edi}
    \mathrm{EDI} = \dfrac{\log{F}-\log{H}}{\log{F}+\log{H}}, \quad H = \dfrac{a}{a+c}, \quad F = \dfrac{b}{b+d},
\end{equation}
where $F$ is the false alarm rate, $H$ is the hit rate, $a$ are the hits, $b$ are the false alarms, $c$ are the misses, and $d$ the correct negatives \citep{Wilks2019}.
Positive values of EDI indicate higher skill than a random forecast. We choose this metric because it is base-rate independent, equitable, and it does not degenerate for rare event classifiers. Thus, the EDI between both thresholds considered can be compared, which is not the case for base-rate dependent measures \citep{Wulff2019}. The equitable threat score is defined as
\begin{equation}\label{eq:ets}
    \mathrm{ETS} = \dfrac{\mathrm{TS}-\mathrm{TS}_\mathrm{ref}}{1-\mathrm{TS}_\mathrm{ref}}, \quad \mathrm{TS} = \dfrac{a}{a+b+c},
\end{equation}
where $\mathrm{TS}_\mathrm{ref}$ is the threat score of a random forecast, and higher ETS is representative of higher skill.

\begin{figure}[h]
 \noindent\includegraphics[width=39pc]{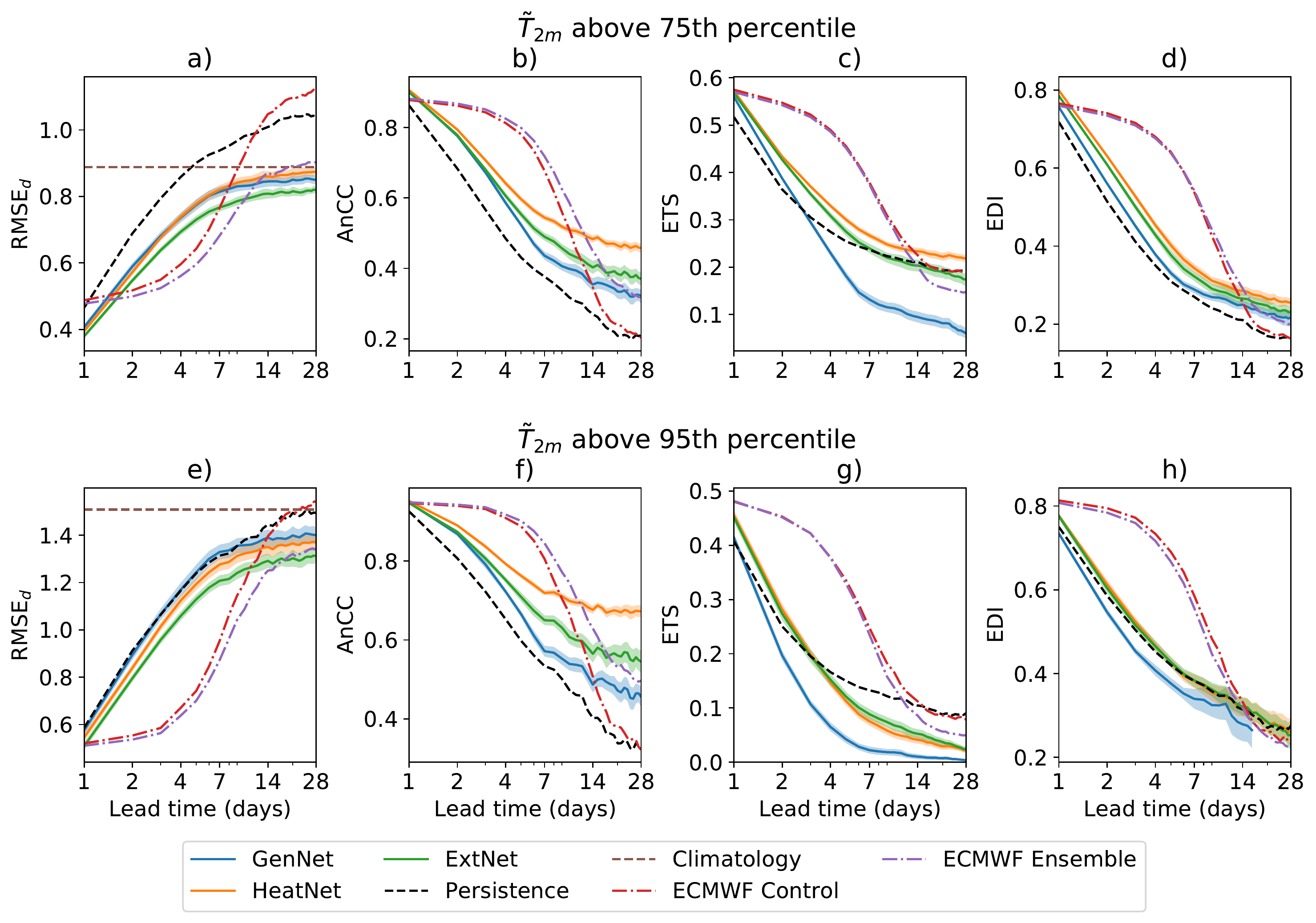}\\
 \caption{Forecast metrics for different models during the summer months of 2017-2021 and over land. Metrics are shown for forecasts conditioned on target standardized temperature anomalies being above the 75th (top) and 95th (bottom) percentiles. From left to right, the debiased root mean squared error (RMSE$_d$), centered anomaly correlation coefficient (AnCC), equitable threat score (ETS) and extremal dependence index (EDI) are shown. Uncertainty bands, shown for the NWMs as a reference, represent 1 standard deviation. Results are only shown for metrics with a robust uncertainty estimate; details may be found in Appendix A.
 }
 \label{fig:metrics_with_lead_time}
\end{figure}

The skill of the different models over land is shown in Figure \ref{fig:metrics_with_lead_time} for the summers of 2017-2021. The lead time is shown in a logarithmic scale to differentiate between three different time scales: the short range ($< 3$ days), the medium range ($3-10$ days), and the extended or subseasonal range ($11-28$ days). In the short range errors are dominated by the initialization, which is more precise for the NWMs, since ERA5 data are fed as predictors. The medium range is characterized by predictable trajectories of the atmospheric state, whereas forecasting a single physical trajectory in the extended range typically adds little value over climatology \citep{LORENZ1969}. Predictive power in the subseasonal range is associated with slower dynamical modes of the climate system, like the MJO or those arising from ocean-atmosphere interactions \citep{Palmer1993, Zhou2019}.

The extreme-focused HeatNet and ExtNet outperform GenNet for both temperature thresholds and all metrics considered, highlighting the usefulness of the exponential loss \eqref{mse_exp} in the extreme prediction task. All NWMs maintain a higher anomaly correlation with the targets than persistence, but only the models trained on the exponential loss improve upon persistence in a mean squared error sense during extremely hot days ($\tilde{T}_{2m} > 95$th percentile). HeatNet, which is trained to emphasize positive extremes exclusively, yields forecasts with higher AnCC than the symmetric ExtNet during hot summer days. Although the AnCC difference indicates higher predictive potential compared to ExtNet in this task, its one-sided emphasis on heatwaves leads to a significant positive bias, as shown in Figures \ref{fig:smoothing_pdf} and \ref{fig:physical_metrics_general}. This bias is detrimental to the prediction of dynamic temperature anomalies, increasing the RMSE$_d$, and does not lead to significant classification skill improvements over ExtNet (Fig. \ref{fig:metrics_with_lead_time}g, h).

The skill of all models is comparable for day-ahead forecasting. In the medium range, the control and ensemble ECMWF forecasts remain superior, but their skill drops significantly faster than that of the NWMs beyond the first week. After the second week, the extreme-focused NWMs have higher regressive skill than the physics-based models. The RMSE$_d$ skill of NWMs relative to the ECMWF system is significantly higher when considering all hot days ($\tilde{T}_{2m} > 75$th percentile) than during extremely hot days ($\tilde{T}_{2m} > 95$th percentile). Positive skill in the extended range is enabled by the use of the exponential loss, without which the NWM forecasts cannot improve upon the AnCC of the operational ensemble system.

The ECMWF ensemble mean forecast substantially improves upon the regressive skill of the control run in the extended range, with RMSE$_d$ and AnCC metrics closer to ExtNet. However, higher regressive skill in the extended range does not translate into higher classification skill, as shown by the ETS and EDI diagnostics. As classifiers, HeatNet and ExtNet have slightly higher skill than the ECMWF models beyond the medium range for hot days ($\tilde{T}_{2m} > 75$th percentile). During extremely hot days ($\tilde{T}_{2m} > 95$th percentile), all NWMs fail to improve upon the ECMWF system as classifiers, although the use of the exponential loss significantly increases the skill of the NWMs for all regression and classification metrics considered.

We also analyzed interhemispheric differences in skill for summer over land using the same thresholds, and found that all models have higher skill in the Northern Hemisphere. The interhemispheric contrast is higher for the NWMs than for the physics-based models; results are included in the Supplemental Material.

\subsection{Smoothing of forecasts with lead time}
\label{subsec:4_smoothing}

\begin{figure}[h]
 \noindent\includegraphics[width=39pc]{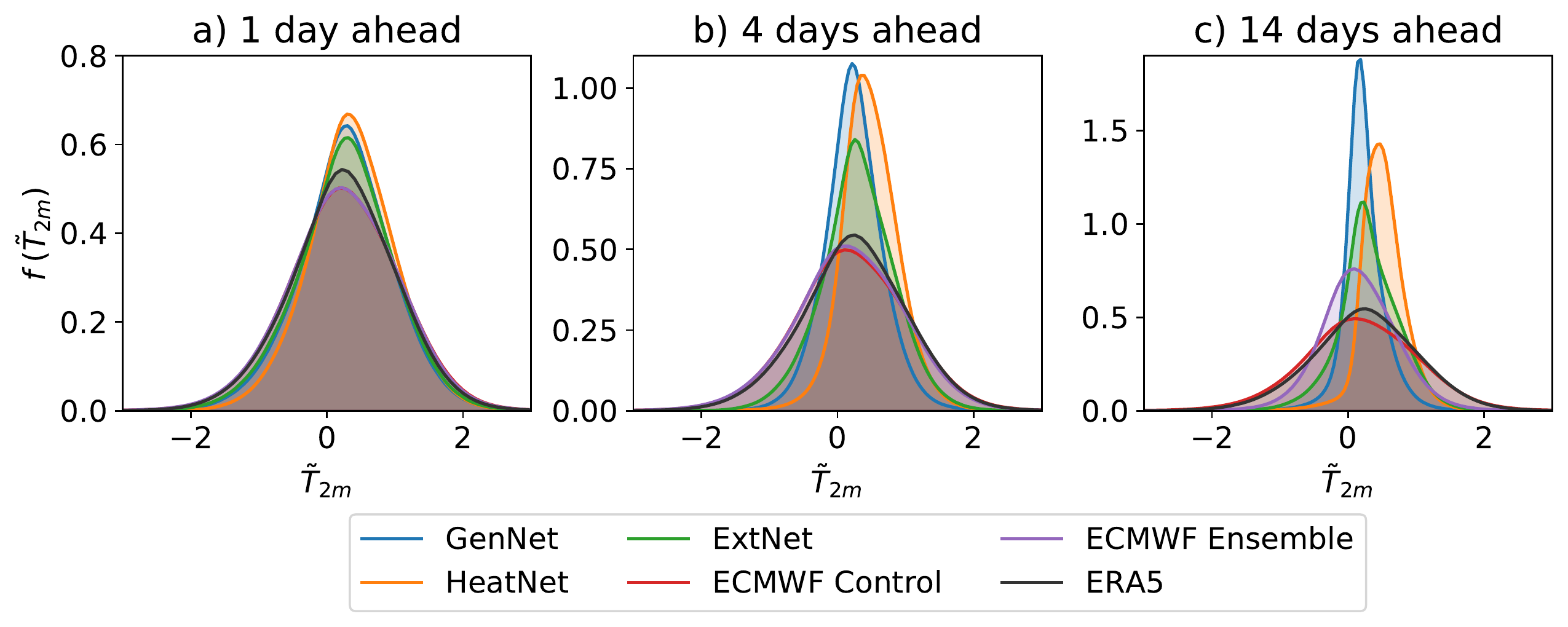}\\
 \caption{Probability density functions (PDFs, here $f(\cdot)$) of forecast global standardized temperature anomalies during the period 2017-2021. Results are shown for all NWMs, the control and ensemble forecasts from the ECMWF S2S system, and the true target distribution (ERA5). Note that the PDFs are not centered about zero, indicating a prediction of the shift in climatology from the 1979-2019 mean.}
 \label{fig:smoothing_pdf}
\end{figure}

The contrast between regressive and classifying skill of NWMs and the ECMWF ensemble is due to a smoothing of their forecasts as lead time progresses, and the predictability of the targets diminishes. Here, we define smoothing as loss of sharpness, or loss of ability to predict events far from climatology. This smoothing is illustrated in Figure \ref{fig:smoothing_pdf} through the evolution of the forecast probability density functions (PDFs) with lead time for all models considered. Smoothing leads to a density concentration near the mean, as the probability of strong temperature anomalies decreases.

In the case of the ECMWF ensemble, lower predictability reduces the correlation between individual forecasts with lead time. This leads to a variance reduction in the ensemble mean distribution. Smoothing is also typical of data-driven methods, although in this case it is the result of forecast error minimization under uncertainty \citep[e.g.,][]{sonderby2020metnet}. While the PDF of individual (hindcast-corrected) physics-based forecasts remains relatively constant, data-driven forecasts shift towards distributions closer to the target mean, with fewer extreme events.

Notably, this smoothing is slowed down through the use of the exponential loss \eqref{mse_exp}, particularly in ExtNet. The use of the symmetric exponential loss increases the probability of significant deviations from climatology: ExtNet forecasts deviations above the 95th target percentile 14 days ahead 4.5 times more frequently than GenNet, and only 25\% less frequently than the ECMWF ensemble mean. Minimizing the positive exponential loss also reduces forecast smoothing, but it leads to a positive bias and makes HeatNet forecasts of negative anomalies extremely unlikely. The deviation of the forecast distribution from the true target PDF is further quantified in Figure \ref{fig:physical_metrics_general} through the Kullback-Leibler (KL) divergence, which is an information-based measure of the difference between probability distributions \citep{Kullback1951, Joyce2011}. The use of the symmetric exponential loss reduces the divergence of ExtNet to less than half of the GenNet divergence for all lead times, whereas the bias induced by the positive extreme loss results in a similar KL divergence compared to GenNet.

Although ExtNet does not manage to capture the same sharpness as the ECMWF ensemble, it is closer in KL divergence to it than to the MSE-trained GenNet model, highlighting the effectiveness of the exponential loss \eqref{mse_exp} in retaining forecast sharpness for a given architecture. Interestingly, the difference in probability of strong positive anomaly forecasts between ExtNet and the ECMWF ensemble mean in the extended range is significantly smaller than the difference in negative anomaly probabilities (Figure \ref{fig:smoothing_pdf}c), even though the loss used to train ExtNet is symmetric. This suggests that positive anomalies are easier to capture than negative anomalies given our predictors.

\subsection{Global surface temperature prediction skill}
\label{subsec:4b_general_skill}

\begin{figure}[h]
 \noindent\includegraphics[width=39pc]{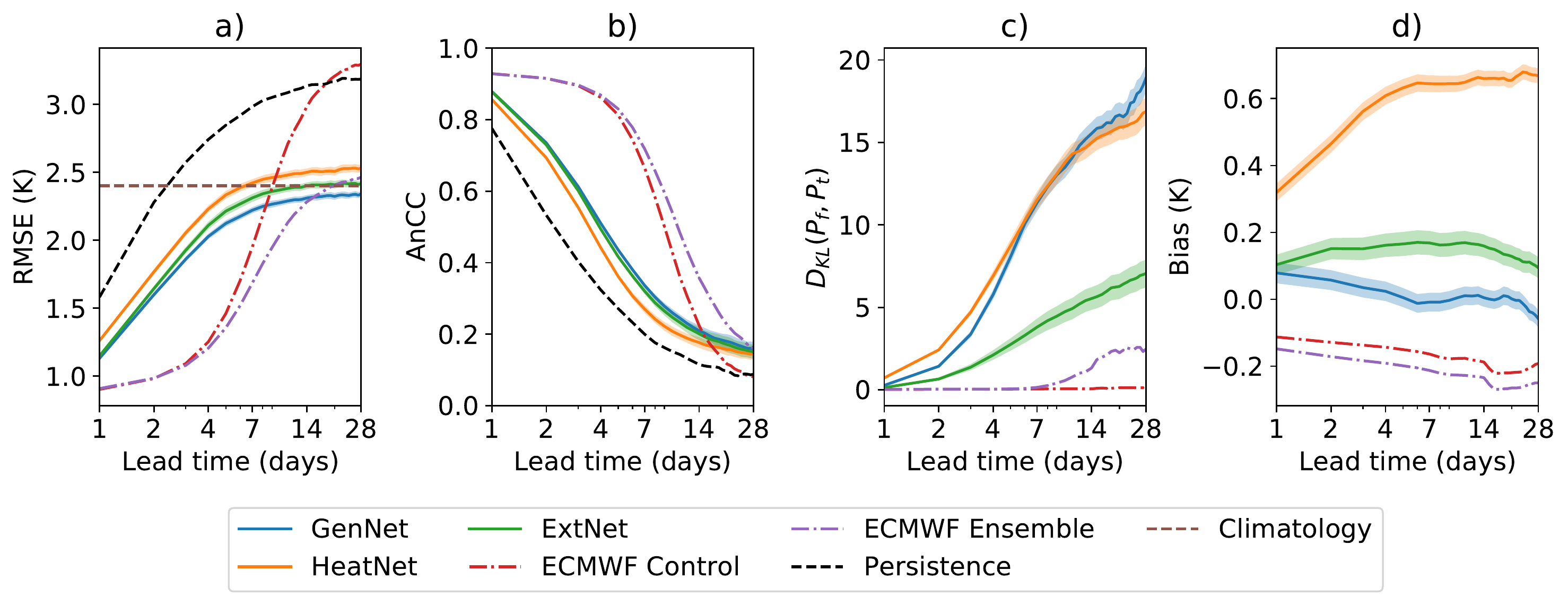}\\
 \caption{Forecast metrics for the general 2 m temperature prediction task ($T_{2m}$), for different models as a function of lead time. From left to right, the root mean squared error (RMSE), the anomaly correlation coefficient (AnCC), the Kullback-Leibler divergence $D_\mathrm{\textit{KL}}$ of $\tilde{T}_{2m}$ with respect to the targets, and the unconditional $T_{2m}$ bias. All results are global and temporal averages over the period 2017-2021. NWM names follow Section \ref{sec:2_problem_statement}\ref{subsec:2b_losses}. Uncertainty defined as in Figure \ref{fig:metrics_with_lead_time}.}
 \label{fig:physical_metrics_general}
\end{figure}

To further assess the effect of the exponential loss \eqref{mse_exp} on the general temperature prediction problem, we include in Figure \ref{fig:physical_metrics_general} the RMSE and AnCC of $T_{2m}$ (i.e., not standardized) for all dates in the test set, over both land and oceans. Note that the RMSE in this case is not debiased. Remarkably, ExtNet shows a very small reduction in forecast skill in the general temperature prediction problem with respect to GenNet. All NWMs beat persistence for all lead times and remain skillfull with respect to the ECMWF control beyond the medium range; the ECMWF ensemble mean remains the most skillful model in the general prediction task. Although the RMSE of ExtNet forecasts converges to that of climatology after three weeks, the model can forecast strong deviations from climatology, as shown in Figure \ref{fig:pnw_heatwave} for an individual forecast at 23 days of lead time. Finally, the forecast biases of GenNet and ExtNet are similar in magnitude to those of the ECMWF model (Fig. \ref{fig:physical_metrics_general}d), even though the neural weather predictions are not bias-corrected by reforecasts. HeatNet does suffer from a significant positive bias, which explains its loss of skill with respect to ExtNet. From Figures \ref{fig:metrics_with_lead_time}--\ref{fig:physical_metrics_general}, it is evident that ExtNet provides the best compromise between extreme heat forecasting skill, forecast reliability and general prediction accuracy among the NWMs considered.

Figure \ref{fig:physical_metrics_general}a also allows comparison with other NWMs in the literature. \cite{Weyn2021} use a neural network with a simpler albeit similar architecture as a time integrator, forecasting fields 6 and 12 hours into the future with each inference step. They show that only when taking an ensemble mean of such models can they beat the RMSE of the ECMWF S2S control forecast in the extended range. In contrast, producing all lead time predictions at once, a single ExtNet forecast is able to improve upon the ECMWF control forecast both in RMSE and AnCC in the extended range. This is consistent with studies comparing direct and iterative forecasting using NWMs, which show that the former configuration leads to enhanced regressive skill \citep{Rasp2020}. The similarities between the ensemble forecast of \cite{Weyn2021}, the ECMWF ensemble, and our results, suggest that NWMs outputting longer lead times yield forecasts more similar to the ensemble mean of physics-driven forecasts than to a given physical trajectory. The similarities include the smoothing of forecasts with lead time, and the saturation of the RMSE in the extended range around the climatological error.

The results in Figures \ref{fig:metrics_with_lead_time}-\ref{fig:physical_metrics_general} yield important insights into the questions that we posed in the introduction. NWMs trained on limited historical data can improve upon persistence in the prediction of out-of-sample rare events, in a regressive sense. As classifiers of extreme events, they only remain skillful with respect to persistence in the short range, due to their loss of sharpness with lead time. For our chosen architecture, positive regressive and classifier skill can only be achieved for extreme events when employing the exponential loss (\ref{mse_exp}). Furthermore, training on the symmetric exponential loss, ExtNet is able to reduce the prediction error for extreme events and slow down the distributional shift with lead time, all while maintaining an unconditional regressive skill practically indistinguishable from models trained on the MSE. Finally, the extreme-focused models improve upon the ECMWF models in the prediction of rate events in a regressive sense after two weeks; in the medium range the physics-based models remain vastly superior. We now explore two specific heatwave events as forecast by the ECMWF model and the NWMs to illustrate the implications of these results.

\subsection{Analysis of the 2017 western European heatwave}
\label{subsec:iberian_heatwave}

Sections \ref{sec:4_results}\ref{subsec:4a_summer_land}--\ref{subsec:4b_general_skill} highlight the different ways in which uncertainty affects physics-based and NWM forecasts. These differences are further explored here at the regional scale by considering the western European heatwave of June 2017. The 2017 heatwave resulted in the hottest June on record in Spain and the Netherlands, and the second warmest in France and Switzerland. It was associated with northward warm air intrusions fostered by a subtropical ridge over western Europe, as shown by \cite{sanchez_benitez_2018}. 

\begin{figure}[h]
 \noindent\includegraphics[width=39pc]{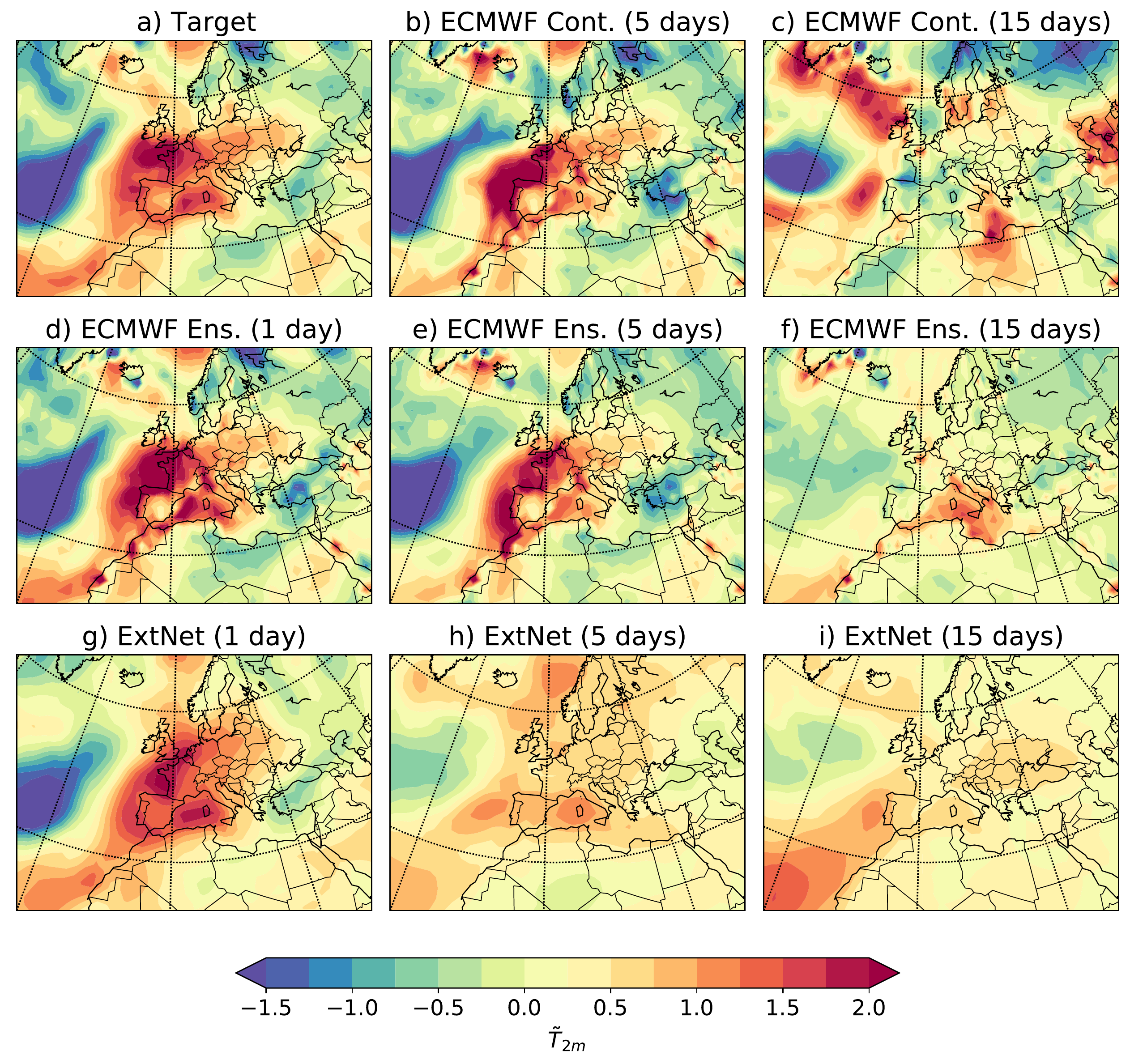}\\
 \caption{Daily-averaged standardized 2-m temperature anomaly over Europe on June 20, 2017 from (a) the ERA5 reanalysis product, and as forecast by the ECMWF S2S control (ECMWF Cont.; b, c), by the ECMWF perturbed ensemble mean (ECMWF Ens.; d, e, f), and by ExtNet (g, h, i). The lead time of the forecasts follows the title of each subfigure.}
 \label{fig:june_iberian_heatwave}
\end{figure}

Forecasts of the standardized temperature anomaly on June 20, 2017 are shown in Figure \ref{fig:june_iberian_heatwave} for several lead times. The ECMWF S2S forecasts initialized 5 days prior accurately predicted the spatial anomaly patterns over western Europe, slightly overpredicting their magnitude over coastal regions and Morocco. In contrast, the control forecast initialized 15 days prior projected important negative temperature anomalies over most of western Europe, opposite of what was observed. It also failed to predict the warm air intrusion from the Saharan coast. Only about 10 of the 50 ECMWF perturbed ensemble members predicted warm temperature anomalies over France and Spain, and a higher fraction predicted negative anomalies; forecasts from 22 of these members are shown in Appendix C for reference. As a result, the ensemble mean forecast was close to climatology outside the Mediterranean Sea.

On the other hand, ExtNet robustly forecast the warm air intrusion for the same lead time (15 days), but not its northward penetration into France and the Benelux. At 5 days lead time, ExtNet predicted positive temperature anomalies over Europe, although the forecast was too mild and inferior to the ECMWF forecasts. Overall, the NWM forecasts track well both the magnitude and patterns of temperature anomaly in the short range. In the medium and extended ranges, the forecasts match the temperature anomaly patterns well, but underestimate their magnitude. Figure \ref{fig:june_iberian_heatwave} is consistent with our hypothesis that, contrary to physics-based models, forecasts by direct NWMs do not represent trajectories of the system. They are more closely related to the mean projection of an ensemble of physics-based forecasts, or NWMs acting as time integrators \citep{Slingo2011, Weyn2021, Scher2021}.

\subsection{Analysis of the 2021 western North American heatwave}
\label{subsec:pnw_heatwave}

\begin{figure}[h]
 \noindent\includegraphics[width=39pc]{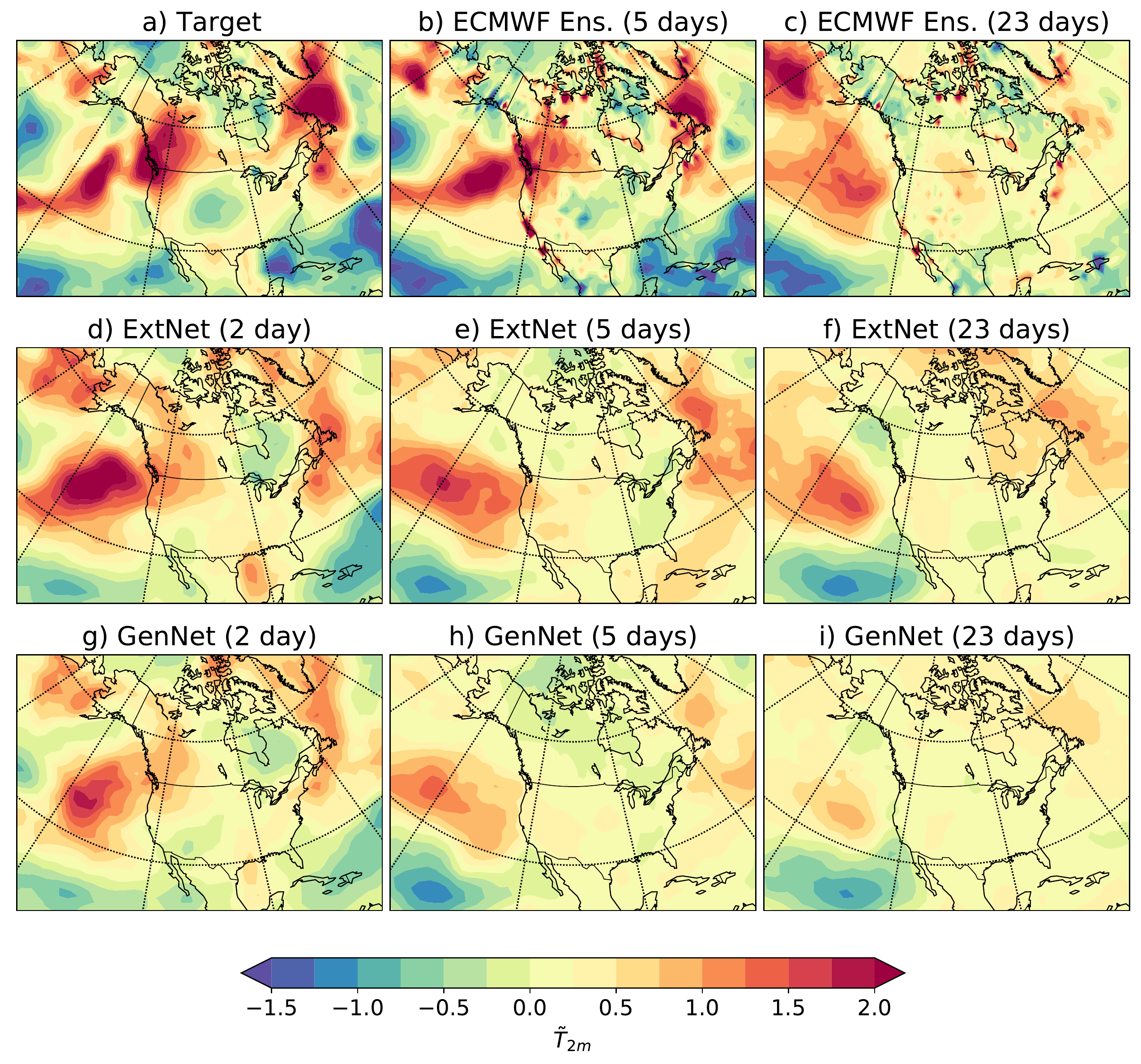}\\
 \caption{Daily-averaged standardized 2-m temperature anomaly over North America on June 26, 2021 from (a) the ERA5 reanalysis product, and as forecast by the ECMWF perturbed ensemble mean (ECMWF Ens.; b, c), ExtNet (d, e, f) and GenNet (g, h, i). The lead time of the forecasts follows the title of each subfigure.}
 \label{fig:pnw_heatwave}
\end{figure}

To showcase the benefits of the symmetric exponential loss, we compare forecasts of the 2021 western North American (WNA) heatwave provided by ExtNet, GenNet and the ECMWF ensemble in Figure \ref{fig:pnw_heatwave}. We consider the WNA heatwave because its forecast using operational systems is well characterized in the literature \citep{Lin2022}. Several phenomena have been suggested as causes of the WNA heatwave. \cite{Lin2022} note the eastward propagation of a Rossby wave train from the tropical western Pacific that may have favored the formation of a heat dome over western North America. \cite{Mo2022} and \cite{Lin2022} also show that the heatwave was preceded by a strong atmospheric river transporting warm moist air from Southeast Asia into the region.

We focus on the heatwave onset, which took place June 25-26, 2021. The actual temperature anomaly on June 26 was characterized by heatwave conditions over Washington, Oregon, and British Columbia. Extreme temperature anomalies were also observed over the northeastern Pacific and the Labrador Sea (Figure \ref{fig:pnw_heatwave}a). The ECMWF ensemble forecast from June 21 correctly predicted warm temperature anomalies over western North America. The forecast from June 3, more than 3 weeks ahead, failed to predict positive anomalies over western North America or the Labrador Sea. This loss of predictive skill over land has been linked to the inability to forecast both the continental penetration of the atmospheric river \citep{Mo2022}, and the eastward shift of the atmospheric ridge over western Canada \citep{Lin2022}.

ExtNet forecast the anomaly pattern correctly with 2 days of lead time, but only predicted significant positive anomalies over Washington, Oregon and the Labrador Sea in the forecast 5 days prior. Compared to GenNet, ExtNet provides significantly better and sharper forecasts for all lead times considered, confirming the results in Sections \ref{sec:4_results}\ref{subsec:4a_summer_land}--\ref{subsec:4b_general_skill}. GenNet underpredicted the extent of the heatwave over North America even in the short range, and failed to predict continental penetration 5 days prior to the event. At 23 days of lead time, the ExtNet forecast closely resembles that of the ECMWF ensemble, exhibiting an anomaly dipole over the eastern Pacific (Figure \ref{fig:pnw_heatwave}c, f). The data-driven model exhibits better correlation with the target over the Labrador and Bering seas at this lead time, highlighting the skill of the model in the extended range.

\section{Model interpretation}
\label{sec:interpretability}

The NWMs presented here leverage a wider range of predictors than other extreme heat forecasting systems in the literature \citep{Chattopadhyay2020, jacques_dumas_2021}. Here we assess the importance of the additional input fields using integrated gradients for feature attribution \citep{Sundararajan2017, rain_check}.

\subsection{Feature attribution through integrated gradients}
\label{sec:IG_method}

The attribution for each feature $x^{(i)}$ is defined as the mean absolute value of its contribution to the model forecast $y'=\Psi(\theta^*,x)$ with respect to a null-contribution baseline forecast $y'_b=\Psi(\theta^*,x_b)$. We formulate the baseline input $x_b$ to be the feature vector on the linear path between $x_\mathrm{min}$ and $x_\mathrm{max}$ that results in a forecast closest to global climatology ($y'\approx0$), where $x_\mathrm{min}$ and $x_\mathrm{max}$ are feature vectors constructed using the minimum and maximum values of the features found in the evaluation set, respectively.

For each actual forecast $y_a'=\Psi(\theta^*,x_a)$, the contribution is computed as the partial derivative of the forecast with respect to $x^{(i)}$, integrated along a linear path from the baseline $x_b=[x^{(1)}_b, x^{(2)}_b, \dots]^T$ to the actual input value $x_a=[x^{(1)}_a, x^{(2)}_a, \dots]^T$,

\begin{equation}\label{eq:ig_att}
    \mathrm{Att}(x^{(i)})\vert_{y'_a} = \dfrac{1}{N}\left\lVert \int_{0}^{1}\dfrac{\delta \Psi(\theta^*, x(\alpha))}{\delta x^{(i)}}d\alpha ~(x_a^{(i)} - x_b^{(i)}) \right\rVert_1,
\end{equation}

where $N$ is the number of pixels over which the L1 norm is computed, $\delta\Psi/\delta x^{(i)}$ is a discretized approximation of the partial derivative, and $\alpha \in [0, 1]$ parameterizes the linear path from baseline to actual feature values, such that $x(\alpha=0)=x_b$ and $x(\alpha=1)=x_a$. Finally, we compute the mean attribution for each feature over dates in the evaluation set.

\subsection{Relevance of model inputs}

\begin{figure}[h]
 \noindent\includegraphics[width=39pc]{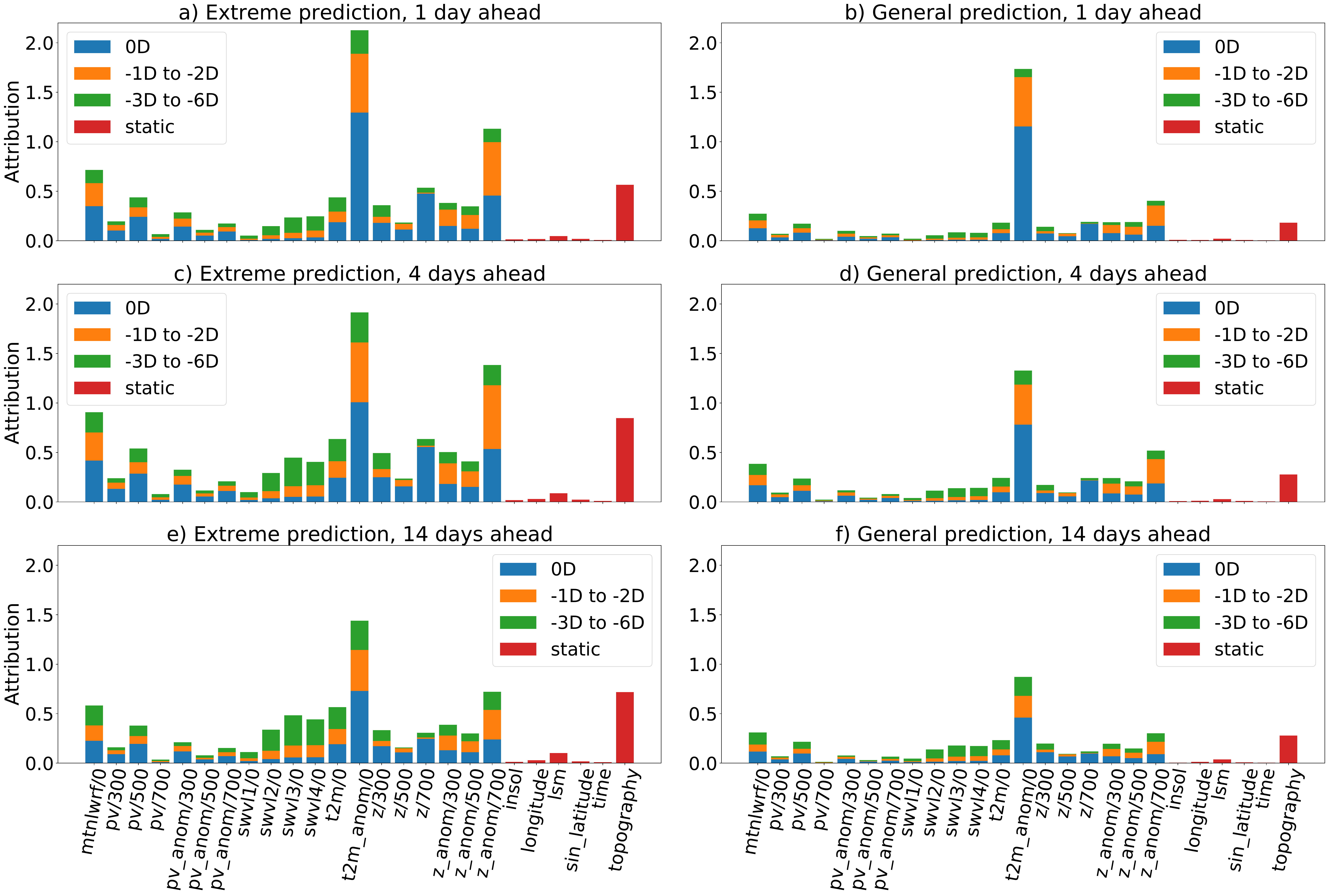}\\
 \caption{Forecast feature attribution for summer over land, using data spanning the period 2015-2019. Attribution is shown for extreme ($\tilde{T}_{2m}$>2) and general prediction, and for different lead times. Feature names follow the notation \textit{name/pressure level}, where each \textit{name} follows ECMWF notation: \textit{mtnlwrf} is outgoing longwave radiation, $pv$ is potential vorticity, $z$ is geopotential, $swvli$ is soil water volume at level $i$ below the ground, and $t2m$ is temperature $2~\mathrm{m}$ above the surface. Pressure level $0$ represents the land surface. The legend (-$iD$) defines the contribution of data $i$ days before forecast initialization.}
 \label{fig:feature_attribution}
\end{figure}

We apply the integrated gradients methodology described in Section \ref{sec:interpretability}\ref{sec:IG_method} to ExtNet forecasts for summer over land during the 5-year period from 2015 to 2019. Feature attributions are shown in Figure \ref{fig:feature_attribution} for the extreme and the general prediction task, and for lead times spanning the short, medium and extended range. The contributions from the most recent data, data from the previous 2 days, and data from the first 4 days of the week preceding the forecast are shown in different colors to quantify the relevance of past history as a predictor of future states.

We find that $\tilde{T}_{2m}$ itself is the most important feature in all cases, while the relative relevance of other features increases with forecast lead time and depends strongly on the task considered (i.e., extreme or general prediction). In addition, the relevance of feature history robustly increases with lead time, suggesting that NWM forecasts in the extended range learn to rely on lower frequency signals. These shifts in relevance with lead time may be used to optimize the NWM architecture, for instance by pruning connections between short lead time forecasts and predictors more than a few days old.

In the extreme prediction task, the temperature anomaly, $700$ hPa geopotential height anomaly, topography and OLR are important predictors at all lead times. Soil moisture below $7~\mathrm{cm}$ gains relevance with lead time, consistent with the characteristic low frequency of land-atmosphere coupling processes and the memory of root-zone soil moisture \citep{Wu2002}. Additional relevant predictors include the potential vorticity at 500 hPa, and the geopotential height anomaly at 300 and 500 hPa. The potential vorticity at 700 hPa and the surface soil moisture above $7~\mathrm{cm}$ are irrelevant to ExtNet predictions at all timescales.

The most important predictors for extreme prediction are also the most important ones for the general forecasting task. However, the relative importance of the temperature anomaly is significantly greater in the general problem, dominating the total attribution. Soil moisture plays a much smaller role in this task compared to heatwave prediction, consistent with observations of a much stronger land-atmosphere coupling under extreme conditions \citep{Orth2012, Benson2021}. Overall, Figure \ref{fig:feature_attribution} suggests that generic estimates of the relevance of alternative features for NWM forecasting may underestimate the contribution of such predictors to extreme event forecasting. Regarding the auxiliary features, we find topography to be the most relevant predictor, followed by the land sea mask. Finally, the global attribution decreases with lead time, consistent with the progressive loss of predictive information and forecast sharpness.

\section{Discussion}
\label{sec:6_discussion}

Referencing the first question that we posed in the introduction, we find that deep learning systems trained on limited historical data can forecast out-of-sample extreme heat events with positive regressive skill above persistence for lead times between 1 and 28 days. This is remarkable given the length of the reanalysis record, and potentially indicative of the ability of regression-based neural weather models to learn about causal physical mechanisms that are common to both the extreme and general forecasting tasks. The rare nature of heatwaves implies that this learning process occurs in the low data regime, and that improved models may be obtained through data augmentation techniques \citep{Miloshevich2022}. In this context, an interesting research direction would be to train deep learning models using a much larger synthetic dataset as a first step \citep{Chattopadhyay2020, jacques_dumas_2021}, and then leverage reanalysis products like ERA5 to fine-tune the model through transfer learning. This technique has already resulted in remarkable achievements in other fields of science, such as organic synthesis \citep{Pesciullesi2020}.

Regarding the second question, we find that NWMs trained on the mean squared error loss fail to yield skillful forecasts of extremely hot days at any lead time considered, at least with our architecture and only using historical data. Our results suggest that it is crucial to train models using losses that emphasize extremes to achieve positive skill in this task, which has been shown before for idealized dynamical systems \citep{Qi_Majda_2020}. Moreover, the switch to the proposed symmetric exponential loss results in negligible skill loss in the general temperature prediction problem, and yields more reliable and sharper forecast distributions further into the future. Thus, the answer to our third question, whether NWMs trained to predict extremes retain skill in more general settings, is positive.

Our best neural weather model (ExtNet) compares favorably to the ECMWF S2S control forecast in the subseasonal range, yielding lower errors and higher correlations with the target both in the general and extreme heat prediction tasks. In the medium range, the ECMWF model remains the most powerful forecast system. The ECMWF ensemble pushes the dominance of physics-based forecasts to longer lead times, but even then ExtNet retains regressive skill in the extreme prediction task after two weeks. This, however, does not fully translate into higher skill as a binary classifier due to the smoothing of forecasts with lead time, which also results in reduced effective resolution. Although the symmetric exponential loss reduces the distributional shift of the forecasts, additional modifications to NWMs, such as the use of generative modeling \citep{Kingma2013, rezende15}, may be necessary to further increase forecast sharpness beyond the short range. This requirement is particularly important for the prediction of extremes. In addition, many practical applications require higher-resolution forecasts than those provided by the neural weather models analyzed here. Higher sharpness and effective resolution at long lead times are some of the specifications that neural weather models will need to meet before they can be used to produce actionable information; we expect the results in this paper to inform the design of such models.

Operational warning systems achieve maximum economic value when they can represent the space of possible trajectories as a probability density function, such that the occurrence of extreme events can be treated probabilistically, not as a binary problem \citep{Palmer2017}. This is done in practice through the use of perturbed ensembles. The use of perturbed ensembles has recently been explored for NWMs acting as time-integrators \citep{Scher2021}, which still show a moderate distributional shift with lead time \citep{Weyn2021}. The use of our proposed exponential loss may enable the use of longer time steps in iterative NWMs while preserving forecast sharpness.

An alternative avenue of research that may prove fruitful is the direct prediction of the probability distribution of trajectories \citep{sonderby2020metnet}, or some parametric approximation of it. In the context of climate modeling, \cite{Guillaumin2021} trained a convolutional neural network to predict the mean and standard deviation of subgrid-scale momentum fluxes in the ocean, which they parameterized as Gaussian. Similar approaches could be taken to predict the ensemble distribution in temperature anomaly projections, retaining the regressive skill of direct NWM forecasts while correcting their underdispersion and smoothness. We hope that these or other methodologies, combined with the use of extreme-focused loss functions such as the one we propose, can enable reliable, actionable and efficient forecasting of extreme events using neural weather models in the near future.




%

%

\clearpage
\acknowledgments The authors thank Stephan Hoyer, Tapio Schneider and John Anderson for valuable discussions that helped improve this paper; as well as Peter D\"{u}ben and two anonymous reviewers for insightful comments on an earlier version of this work. The authors also acknowledge the use of the DLWP-CS open source package developed by Jonathan Weyn as a starting point for this project.


%
%
\datastatement The ERA5 reanalysis data is freely available at the Copernicus Climate Data Store (\url{https://cds.climate.copernicus.eu}). The ECMWF S2S control and perturbed forecasts can be obtained from the Meteorological Archival and Retrieval System of ECMWF (\url{https://apps.ecmwf.int/datasets/data/s2s}). The software used to train the neural weather models is available on GitHub (\url{https://github.com/google-research/heatnet}).


%






%



\appendix[A]
\appendixtitle{Metric uncertainty estimation}

The test set used in this article contains about 7 million samples,  $>170,000$ samples of hot days over land ($\tilde{T}_{2m} > 75$th percentile), and $>34,000$ samples of extremely hot days over land ($\tilde{T}_{2m} > 95$th percentile). In order to determine the variance of the sample mean due to finite sample size, we use block bootstrapping with year-long disjoint blocks \citep{Hall1995}. We construct the empirical distribution function of the sample mean from $10^6$ bootstrap samples, and use its standard deviation as a measure of uncertainty. All uncertainty estimates proved robust to block size reduction except for EDI during extremely hot days and after a certain lead time for GenNet. Results are omitted for these EDI estimates.

\appendix[B]\appendixtitle{Regressive skill conditioned on target and forecast values}

\begin{figure}[h]
 \noindent\includegraphics[width=19pc]{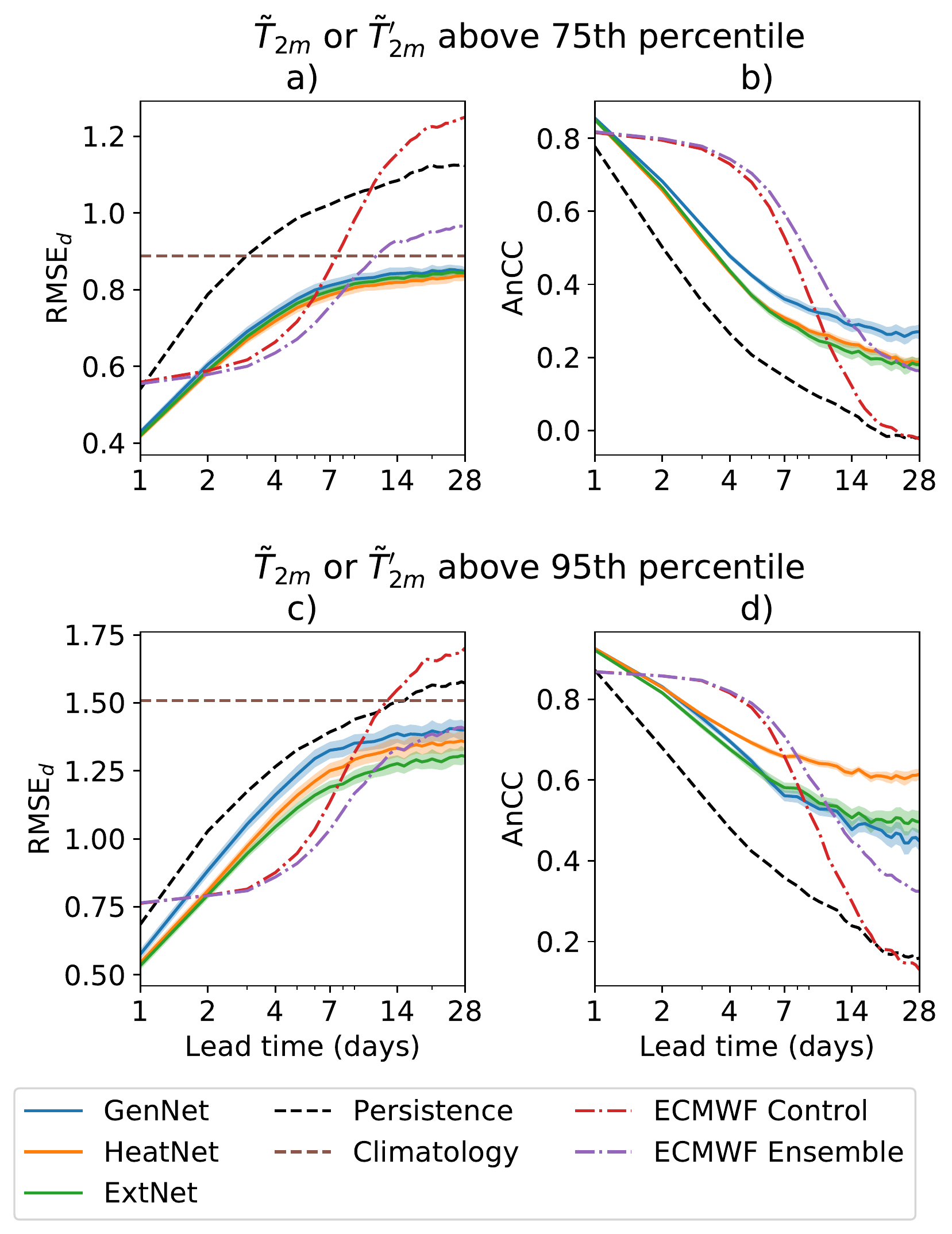}\\
 \caption{Forecast metrics for different models during the summer months of 2017-2021 and over land. Metrics are shown for forecasts conditioned on either target or forecast standardized temperature anomalies being above the 75th (top) and 95th (bottom) percentiles of values in the test set. Legend as in Figure \ref{fig:metrics_with_lead_time}.
 }
 \label{fig:metrics_with_lead_time_joint}
\end{figure}

To verify that the evaluated models do not suffer from the forecaster's dilemma \citep{Lerch2017}, the debiased RMSE and the centered anomaly correlation coefficient are evaluated here over all dates and locations where either the target or the forecast temperature anomalies were above a certain percentile of values in the test set. This conditioning assesses the skill over false alarms, as well as over hits and misses, penalizing models that overforecast extremes. As shown in Figure \ref{fig:metrics_with_lead_time_joint}, differences with respect to Figure \ref{fig:metrics_with_lead_time} are most prominent for the ECMWF and persistence forecasts, which are well calibrated. The skill reduction is smaller for NWMs, which tend to underpredict extremes, and insignificant for GenNet. This pushes the threshold above which ExtNet improves upon GenNet to a higher percentile. Above the 95th percentile, ExtNet still outperforms GenNet.

\appendix[C]
\appendixtitle{Stamp plot of the 2017 European heatwave from the ECMWF ensemble}

\begin{figure}[h]
 \noindent\includegraphics[width=39pc]{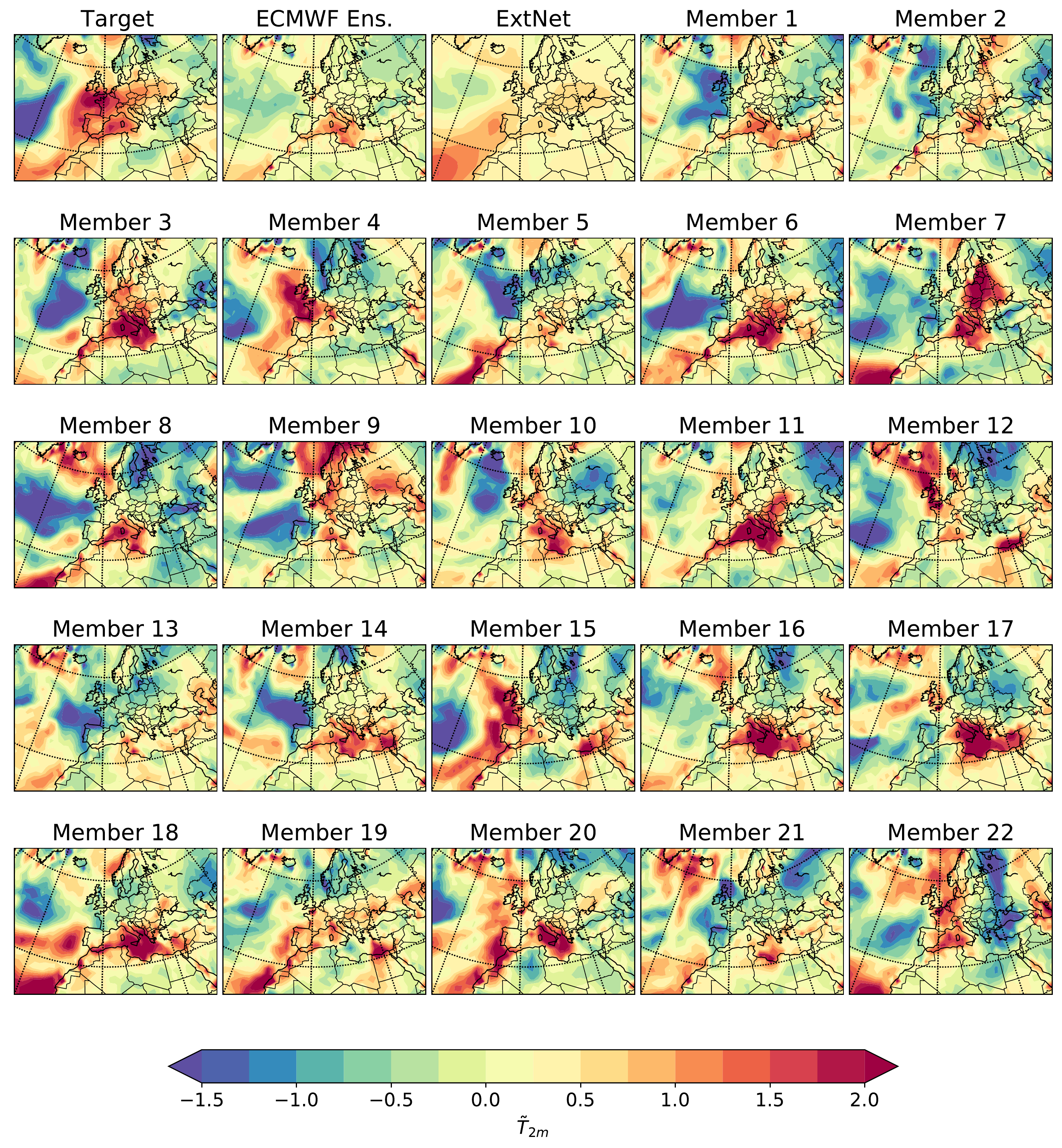}\\
 \caption{Daily-averaged standardized 2-m temperature anomaly over Europe on June 20, 2017 from the ERA5 reanalysis product (Target), the ECMWF ensemble mean forecast (ECMWF Ens.), the ExtNet forecast, and the forecast by 22 individual members of the operational ECMWF perturbed ensemble. The lead time for all forecasts is 15 days.
 }
 \label{fig:stamp_plots_Iberian_heatwave}
\end{figure}

Figure \ref{fig:stamp_plots_Iberian_heatwave} shows a stamp plot of 22 random individual 15-day forecasts of the heatwave described in Section \ref{sec:4_results}\ref{subsec:iberian_heatwave}, from the ECMWF ensemble. Several members (e.g., 4, 15, 20) capture elements of the heatwave, but many others show similar shortcomings to the control forecast.
\clearpage
\clearpage
\bibliographystyle{ametsocV6}
\bibliography{references}

\begin{thebibliography}{84}
\providecommand{\natexlab}[1]{#1}
\providecommand{\url}[1]{\texttt{#1}}
\renewcommand{\UrlFont}{\rmfamily}
\providecommand{\urlprefix}{URL }
\expandafter\ifx\csname urlstyle\endcsname\relax
  \providecommand{\doi}[1]{https://doi.org/\discretionary{}{}{}#1}\else
  \providecommand{\doi}{https://doi.org/\discretionary{}{}{}\begingroup
  \urlstyle{rm}\Url}\fi
\providecommand{\eprint}[2][]{\url{#2}}

\bibitem[{Accadia et~al.(2003)Accadia, Mariani, Casaioli, Lavagnini,, and
  Speranza}]{Accadia2003}
Accadia, C., S.~Mariani, M.~Casaioli, A.~Lavagnini, and A.~Speranza, 2003:
  Sensitivity of precipitation forecast skill scores to bilinear interpolation
  and a simple nearest-neighbor average method on high-resolution verification
  grids. \textit{Weather and Forecasting}, \textbf{18~(5)}, 918--932,
  \doi{10.1175/1520-0434(2003)018<0918:SOPFSS>2.0.CO;2}.

\bibitem[{Benson and Dirmeyer(2021)Benson, and Dirmeyer}]{Benson2021}
Benson, D.~O., and P.~A. Dirmeyer, 2021: Characterizing the relationship
  between temperature and soil moisture extremes and their role in the
  exacerbation of heat waves over the contiguous united states. \textit{J.
  Clim.}, \textbf{34~(6)}, 2175--2187, \doi{10.1175/JCLI-D-20-0440.1}.

\bibitem[{Br{\'{a}}s et~al.(2021)Br{\'{a}}s, Seixas, Carvalhais,, and
  Jägermeyr}]{Bras_2021}
Br{\'{a}}s, T.~A., J.~Seixas, N.~Carvalhais, and J.~Jägermeyr, 2021: Severity
  of drought and heatwave crop losses tripled over the last five decades in
  europe. \textit{Environmental Research Letters}, \textbf{16~(6)}, 065\,012,
  \doi{10.1088/1748-9326/abf004}.

\bibitem[{Buizza et~al.(1999)Buizza, Milleer,, and Palmer}]{Buizza1999}
Buizza, R., M.~Milleer, and T.~N. Palmer, 1999: Stochastic representation of
  model uncertainties in the {ECMWF} ensemble prediction system.
  \textit{Quarterly Journal of the Royal Meteorological Society}, \textbf{125},
  2887--2908, \doi{10.1002/qj.49712556006}.

\bibitem[{Chan et~al.(2020)Chan, Cobb, Zeppetello, Battisti,, and
  Huybers}]{chan2020}
Chan, D., A.~Cobb, L.~R.~V. Zeppetello, D.~S. Battisti, and P.~Huybers, 2020:
  Summertime temperature variability increases with local warming in
  midlatitude regions. \textit{Geophysical Research Letters}, \textbf{47},
  e2020GL087\,624, \doi{10.1029/2020GL087624}.

\bibitem[{Chattopadhyay et~al.(2020)Chattopadhyay, Nabizadeh,, and
  Hassanzadeh}]{Chattopadhyay2020}
Chattopadhyay, A., E.~Nabizadeh, and P.~Hassanzadeh, 2020: Analog forecasting
  of extreme-causing weather patterns using deep learning. \textit{Journal of
  Advances in Modeling Earth Systems}, \textbf{12}, e2019MS001\,958,
  \doi{10.1029/2019MS001958}.

\bibitem[{Deng et~al.(2018)Deng, Ting, Yang,, and Tan}]{Deng2018}
Deng, K., M.~Ting, S.~Yang, and Y.~Tan, 2018: Increased frequency of summer
  extreme heat waves over {Texas} area tied to the amplification of pacific
  zonal {SST} gradient. \textit{J. Clim.}, \textbf{31~(14)}, 5629--5647,
  \doi{10.1175/JCLI-D-17-0554.1}.

\bibitem[{ECMWF(2016)}]{p17118}
ECMWF, 2016: \textit{IFS Documentation CY43R1 - Part V: Ensemble Prediction
  System}. No.~5, IFS Documentation, ECMWF, \doi{10.21957/6fm80smm}.

\bibitem[{ECMWF(2019)}]{p19309}
ECMWF, 2019: \textit{IFS Documentation CY46R1 - Part V: Ensemble Prediction
  System}. No.~5, IFS Documentation, \doi{10.21957/38yug0cev}.

\bibitem[{ECMWF(2020)}]{p19749}
ECMWF, 2020: \textit{IFS Documentation CY47R1 - Part V: Ensemble Prediction
  System}. No.~5, IFS Documentation, ECMWF, \doi{10.21957/d7e3hrb}.

\bibitem[{Espeholt et~al.(2022)}]{Espeholt2022}
Espeholt, L., and Coauthors, 2022: Deep learning for twelve hour precipitation
  forecasts. \textit{Nature Communications}, \textbf{13}, 5145,
  \doi{10.1038/s41467-022-32483-x}.

\bibitem[{Ferro and Stephenson(2011)Ferro, and Stephenson}]{edi_ferro2011}
Ferro, C. A.~T., and D.~B. Stephenson, 2011: Extremal dependence indices:
  Improved verification measures for deterministic forecasts of rare binary
  events. \textit{Weather and Forecasting}, \textbf{26~(5)}, 699 -- 713,
  \doi{10.1175/WAF-D-10-05030.1}.

\bibitem[{Guillaumin and Zanna(2021)Guillaumin, and Zanna}]{Guillaumin2021}
Guillaumin, A.~P., and L.~Zanna, 2021: Stochastic-deep learning
  parameterization of ocean momentum forcing. \textit{Journal of Advances in
  Modeling Earth Systems}, \textbf{13}, e2021MS002\,534,
  \doi{10.1029/2021MS002534}.

\bibitem[{Hall et~al.(1995)Hall, Horowitz,, and Jing}]{Hall1995}
Hall, P., J.~L. Horowitz, and B.-Y. Jing, 1995: On blocking rules for the
  bootstrap with dependent data. \textit{Biometrika}, \textbf{82}, 561--574,
  \doi{10.2307/2337534}.

\bibitem[{He et~al.(2015)He, Zhang, Ren,, and Sun}]{He2015}
He, K., X.~Zhang, S.~Ren, and J.~Sun, 2015: Delving deep into rectifiers:
  Surpassing human-level performance on {ImageNet} classification.
  \doi{10.48550/arXiv.1502.01852}.

\bibitem[{Heo et~al.(2019)Heo, Bell,, and Lee}]{Heo2019}
Heo, S., M.~L. Bell, and J.-T. Lee, 2019: Comparison of health risks by heat
  wave definition: Applicability of wet-bulb globe temperature for heat wave
  criteria. \textit{Environmental Research}, \textbf{168}, 158--170,
  \doi{10.1016/j.envres.2018.09.032}.

\bibitem[{Hersbach et~al.(2020)}]{era5}
Hersbach, H., and Coauthors, 2020: The {ERA5} global reanalysis.
  \textit{Quarterly Journal of the Royal Meteorological Society},
  \textbf{146~(730)}, 1999--2049, \doi{10.1002/qj.3803}.

\bibitem[{Huang et~al.(2020)}]{Huang2020}
Huang, H., and Coauthors, 2020: {UNet} 3+: A {Full-Scale} connected {UNet} for
  medical image segmentation. \doi{10.48550/arXiv.2004.08790},
  \eprint{2004.08790}.

\bibitem[{Inubushi and Goto(2020)Inubushi, and Goto}]{Inubushi2020}
Inubushi, M., and S.~Goto, 2020: Transfer learning for nonlinear dynamics and
  its application to fluid turbulence. \textit{Physical Review E},
  \textbf{102}, 43\,301, \doi{10.1103/PhysRevE.102.043301}.

\bibitem[{Jacques-Coper et~al.(2015)Jacques-Coper, Brönnimann, Martius, Vera,,
  and Cerne}]{jacques_coper2015}
Jacques-Coper, M., S.~Brönnimann, O.~Martius, C.~S. Vera, and S.~B. Cerne,
  2015: Evidence for a modulation of the intraseasonal summer temperature in
  eastern patagonia by the {Madden-Julian} oscillation. \textit{Journal of
  Geophysical Research: Atmospheres}, \textbf{120}, 7340--7357,
  \doi{10.1002/2014JD022924}.

\bibitem[{Jacques-Dumas et~al.(2022)Jacques-Dumas, Ragone, Borgnat, Abry,, and
  Bouchet}]{jacques_dumas_2021}
Jacques-Dumas, V., F.~Ragone, P.~Borgnat, P.~Abry, and F.~Bouchet, 2022: Deep
  learning-based extreme heatwave forecast. \textit{Frontiers in Climate},
  \textbf{4}, \doi{10.3389/fclim.2022.789641}.

\bibitem[{Joyce(2011)}]{Joyce2011}
Joyce, J.~M., 2011: \textit{Kullback-Leibler Divergence}, 720--722. Springer
  Berlin Heidelberg, Berlin, Heidelberg, \doi{10.1007/978-3-642-04898-2_327}.

\bibitem[{Ke et~al.(2016)Ke, Wu, Rice, Kintner-Meyer,, and Lu}]{ke_2016}
Ke, X., D.~Wu, J.~Rice, M.~Kintner-Meyer, and N.~Lu, 2016: Quantifying impacts
  of heat waves on power grid operation. \textit{Applied Energy}, \textbf{183},
  504--512, \doi{10.1016/j.apenergy.2016.08.188}.

\bibitem[{Keisler(2022)}]{Keisler2022}
Keisler, R., 2022: Forecasting global weather with graph neural networks.
  \doi{10.48550/arxiv.2202.07575}.

\bibitem[{Kingma and Ba(2017)Kingma, and Ba}]{kingma2017adam}
Kingma, D.~P., and J.~Ba, 2017: Adam: A method for stochastic optimization.
  \doi{10.48550/arXiv.1412.6980}, \eprint{1412.6980}.

\bibitem[{Kingma and Welling(2013)Kingma, and Welling}]{Kingma2013}
Kingma, D.~P., and M.~Welling, 2013: Auto-encoding variational {Bayes}.
  \doi{10.48550/arxiv.1312.6114}.

\bibitem[{Kullback and Leibler(1951)Kullback, and Leibler}]{Kullback1951}
Kullback, S., and R.~A. Leibler, 1951: On information and sufficiency.
  \textit{The Annals of Mathematical Statistics}, \textbf{22}, 79--86,
  \doi{10.1214/aoms/1177729694}.

\bibitem[{Lerch et~al.(2017)Lerch, Thorarinsdottir, Ravazzolo,, and
  Gneiting}]{Lerch2017}
Lerch, S., T.~L. Thorarinsdottir, F.~Ravazzolo, and T.~Gneiting, 2017:
  Forecaster’s dilemma: Extreme events and forecast evaluation.
  \textit{Statistical Science}, \textbf{32}, 106--127, \doi{10.1214/16-STS588}.

\bibitem[{Li et~al.(2018)Li, Grandvalet,, and Davoine}]{li_2018}
Li, X., Y.~Grandvalet, and F.~Davoine, 2018: Explicit inductive bias for
  transfer learning with convolutional networks. \textit{Proceedings of the
  35th International Conference on Machine Learning}, J.~Dy, and A.~Krause,
  Eds., PMLR, Proceedings of Machine Learning Research, Vol.~80, 2825--2834,
  \urlprefix\url{https://proceedings.mlr.press/v80/li18a.html}.

\bibitem[{Lin et~al.(2022)Lin, Mo,, and Vitart}]{Lin2022}
Lin, H., R.~Mo, and F.~Vitart, 2022: The 2021 western {North American} heatwave
  and its subseasonal predictions. \textit{Geophysical Research Letters},
  \textbf{49}, e2021GL097\,036, \doi{10.1029/2021GL097036}.

\bibitem[{Lopez-Gomez et~al.(2022)Lopez-Gomez, Christopoulos, Ervik, Dunbar,
  Cohen,, and Schneider}]{LopezGomez2022}
Lopez-Gomez, I., C.~Christopoulos, H.~L.~L. Ervik, O.~R.~A. Dunbar, Y.~Cohen,
  and T.~Schneider, 2022: Training physics-based machine-learning
  parameterizations with gradient-free ensemble kalman methods. \textit{Journal
  of Advances in Modeling Earth Systems}, \textbf{14}, e2022MS003\,105,
  \doi{10.1029/2022MS003105}.

\bibitem[{Lopez-Gomez et~al.(2020)Lopez-Gomez, Cohen, He, Jaruga,, and
  Schneider}]{LopezGomez2020}
Lopez-Gomez, I., Y.~Cohen, J.~He, A.~Jaruga, and T.~Schneider, 2020: A
  generalized mixing length closure for eddy-diffusivity mass-flux schemes of
  turbulence and convection. \textit{Journal of Advances in Modeling Earth
  Systems}, \textbf{12}, e2020MS002\,161, \doi{10.1029/2020MS002161}.

\bibitem[{Lorenz(1969{\natexlab{a}})}]{Lorenz1969b}
Lorenz, E.~N., 1969{\natexlab{a}}: Atmospheric predictability as revealed by
  naturally occurring analogues. \textit{Journal of Atmospheric Sciences},
  \textbf{26}, 636--646, \doi{10.1175/1520-0469(1969)26<636:APARBN>2.0.CO;2}.

\bibitem[{Lorenz(1969{\natexlab{b}})}]{LORENZ1969}
Lorenz, E.~N., 1969{\natexlab{b}}: {The predictability of a flow which
  possesses many scales of motion}. \textit{Tellus}, \textbf{21~(3)}, 289--307,
  \doi{10.1111/j.2153-3490.1969.tb00444.x}.

\bibitem[{Malardel et~al.(2016)Malardel, Wedi, Deconinck, Diamantakis,
  Kuehnlein, Mozdzynski, Hamrud,, and Smolarkiewicz}]{ifs_grid}
Malardel, S., N.~Wedi, W.~Deconinck, M.~Diamantakis, C.~Kuehnlein,
  G.~Mozdzynski, M.~Hamrud, and P.~Smolarkiewicz, 2016: A new grid for the
  {IFS}. \textit{ECMWF Newsletter}, 23--28, \doi{10.21957/zwdu9u5i}.

\bibitem[{Maloney et~al.(2019)Maloney, Ángel F~Adames,, and Bui}]{Maloney2019}
Maloney, E.~D., Ángel F~Adames, and H.~X. Bui, 2019: Madden–{Julian}
  oscillation changes under anthropogenic warming. \textit{Nature Climate
  Change}, \textbf{9}, 26--33, \doi{10.1038/s41558-018-0331-6}.

\bibitem[{Mann et~al.(2018)Mann, Rahmstorf, Kornhuber, Steinman, Miller,
  Petri,, and Coumou}]{Mann2018}
Mann, M.~E., S.~Rahmstorf, K.~Kornhuber, B.~A. Steinman, S.~K. Miller,
  S.~Petri, and D.~Coumou, 2018: Projected changes in persistent extreme summer
  weather events: The role of quasi-resonant amplification. \textit{Science
  advances}, \textbf{4}, eaat3272--eaat3272, \doi{10.1126/sciadv.aat3272}.

\bibitem[{Mass et~al.(2002)Mass, Ovens, Westrick,, and Colle}]{Mass2002}
Mass, C.~F., D.~Ovens, K.~Westrick, and B.~A. Colle, 2002: Does increasing
  horizontal resolution produce more skillful forecasts?: The results of two
  years of real-time numerical weather prediction over the pacific northwest.
  \textit{Bulletin of the American Meteorological Society}, \textbf{83},
  407--430, \doi{10.1175/1520-0477(2002)083<0407:DIHRPM>2.3.CO;2}.

\bibitem[{Miller et~al.(2021)Miller, Wang, Li, Harnos,, and Ford}]{Miller2021}
Miller, D.~E., Z.~Wang, B.~Li, D.~S. Harnos, and T.~Ford, 2021: Skillful
  subseasonal prediction of {U.S}. extreme warm days and standardized
  precipitation index in boreal summer. \textit{J. Clim.}, \textbf{34~(14)},
  5887--5898, \doi{10.1175/JCLI-D-20-0878.1}.

\bibitem[{Miloshevich et~al.(2022)Miloshevich, Cozian, Abry, Borgnat,, and
  Bouchet}]{Miloshevich2022}
Miloshevich, G., B.~Cozian, P.~Abry, P.~Borgnat, and F.~Bouchet, 2022:
  Probabilistic forecasts of extreme heatwaves using convolutional neural
  networks in a regime of lack of data. \doi{10.48550/arxiv.2208.00971}.

\bibitem[{Mo et~al.(2022)Mo, Lin,, and Vitart}]{Mo2022}
Mo, R., H.~Lin, and F.~Vitart, 2022: An anomalous warm-season trans-{Pacific}
  atmospheric river linked to the 2021 western {North America} heatwave.
  \textit{Communications Earth \& Environment}, \textbf{3}, 127,
  \doi{10.1038/s43247-022-00459-w}.

\bibitem[{Molteni et~al.(1996)Molteni, Buizza, Palmer,, and
  Petroliagis}]{Molteni1996}
Molteni, F., R.~Buizza, T.~N. Palmer, and T.~Petroliagis, 1996: The {ECMWF}
  ensemble prediction system: Methodology and validation. \textit{Quarterly
  Journal of the Royal Meteorological Society}, \textbf{122}, 73--119,
  \doi{https://doi.org/10.1002/qj.49712252905}.

\bibitem[{Ogi et~al.(2003)Ogi, Tachibana,, and Yamazaki}]{Ogi2003}
Ogi, M., Y.~Tachibana, and K.~Yamazaki, 2003: Impact of the wintertime north
  {Atlantic} oscillation ({NAO}) on the summertime atmospheric circulation.
  \textit{Geophysical Research Letters}, \textbf{30},
  \doi{10.1029/2003GL017280}.

\bibitem[{Orth and Seneviratne(2012)Orth, and Seneviratne}]{Orth2012}
Orth, R., and S.~I. Seneviratne, 2012: {Analysis of soil moisture memory from
  observations in Europe}. \textit{Journal of Geophysical Research:
  Atmospheres}, \textbf{117}, \doi{10.1029/2011JD017366}.

\bibitem[{Palmer(2017)}]{Palmer2017}
Palmer, T., 2017: {The primacy of doubt: Evolution of numerical weather
  prediction from determinism to probability}. \textit{Journal of Advances in
  Modeling Earth Systems}, \textbf{9~(2)}, 730--734,
  \doi{10.1002/2017MS000999}.

\bibitem[{Palmer(1993)}]{Palmer1993}
Palmer, T.~N., 1993: Extended-range atmospheric prediction and the {Lorenz}
  model. \textit{Bulletin of the American Meteorological Society}, \textbf{74},
  49--66, \doi{10.1175/1520-0477(1993)074<0049:ERAPAT>2.0.CO;2}.

\bibitem[{Parente et~al.(2018)Parente, Pereira, Amraoui,, and
  Fischer}]{parente_2018}
Parente, J., M.~Pereira, M.~Amraoui, and E.~Fischer, 2018: Heat waves in
  {Portugal}: Current regime, changes in future climate and impacts on extreme
  wildfires. \textit{Science of The Total Environment}, \textbf{631-632},
  534--549, \doi{10.1016/j.scitotenv.2018.03.044}.

\bibitem[{Pathak et~al.(2022)}]{Pathak2022}
Pathak, J., and Coauthors, 2022: {FourCastNet}: A global data-driven
  high-resolution weather model using adaptive fourier neural operators.
  \doi{10.48550/arxiv.2202.11214}.

\bibitem[{Perkins-Kirkpatrick and Lewis(2020)Perkins-Kirkpatrick, and
  Lewis}]{perkins_2020}
Perkins-Kirkpatrick, S.~E., and S.~C. Lewis, 2020: Increasing trends in
  regional heatwaves. \textit{Nature Communications}, \textbf{11}, 3357,
  \doi{10.1038/s41467-020-16970-7}.

\bibitem[{Pesciullesi et~al.(2020)Pesciullesi, Schwaller, Laino,, and
  Reymond}]{Pesciullesi2020}
Pesciullesi, G., P.~Schwaller, T.~Laino, and J.-L. Reymond, 2020: Transfer
  learning enables the molecular transformer to predict regio- and
  stereoselective reactions on carbohydrates. \textit{Nature Communications},
  \textbf{11}, 4874, \doi{10.1038/s41467-020-18671-7}.

\bibitem[{Qi and Majda(2020)Qi, and Majda}]{Qi_Majda_2020}
Qi, D., and A.~J. Majda, 2020: Using machine learning to predict extreme events
  in complex systems. \textit{Proceedings of the National Academy of Sciences},
  \textbf{117~(1)}, 52--59, \doi{10.1073/pnas.1917285117}.

\bibitem[{Rasp et~al.(2020)Rasp, Dueben, Scher, Weyn, Mouatadid,, and
  Thuerey}]{Rasp2020}
Rasp, S., P.~D. Dueben, S.~Scher, J.~A. Weyn, S.~Mouatadid, and N.~Thuerey,
  2020: {WeatherBench}: A benchmark data set for data-driven weather
  forecasting. \textit{Journal of Advances in Modeling Earth Systems},
  \textbf{12}, e2020MS002\,203, \doi{10.1029/2020MS002203}.

\bibitem[{Ravuri et~al.(2021)}]{Ravuri2021}
Ravuri, S., and Coauthors, 2021: Skilful precipitation nowcasting using deep
  generative models of radar. \textit{Nature}, \textbf{597}, 672--677,
  \doi{10.1038/s41586-021-03854-z}.

\bibitem[{Rezende and Mohamed(2015)Rezende, and Mohamed}]{rezende15}
Rezende, D., and S.~Mohamed, 2015: Variational inference with normalizing
  flows. \textit{Proceedings of the 32nd International Conference on Machine
  Learning}, F.~Bach, and D.~Blei, Eds., PMLR, Lille, France, Proceedings of
  Machine Learning Research, Vol.~37, 1530--1538,
  \urlprefix\url{https://proceedings.mlr.press/v37/rezende15.html}.

\bibitem[{Richardson(2000)}]{Richardson2000}
Richardson, D.~S., 2000: Skill and relative economic value of the {ECMWF}
  ensemble prediction system. \textit{Quarterly Journal of the Royal
  Meteorological Society}, \textbf{126}, 649--667,
  \doi{10.1002/qj.49712656313}.

\bibitem[{Robine et~al.(2008)Robine, Cheung, {Le Roy}, {Van Oyen}, Griffiths,
  Michel,, and Herrmann}]{robine_2008}
Robine, J.-M., S.~L.~K. Cheung, S.~{Le Roy}, H.~{Van Oyen}, C.~Griffiths, J.-P.
  Michel, and F.~R. Herrmann, 2008: Death toll exceeded 70,000 in {Europe}
  during the summer of 2003. \textit{Comptes Rendus Biologies},
  \textbf{331~(2)}, 171--178, \doi{10.1016/j.crvi.2007.12.001}.

\bibitem[{Ronchi et~al.(1996)Ronchi, Iacono,, and Paolucci}]{Ronchi1996}
Ronchi, C., R.~Iacono, and P.~S. Paolucci, 1996: The “cubed sphere”: A new
  method for the solution of partial differential equations in spherical
  geometry. \textit{Journal of Computational Physics}, \textbf{124}, 93--114,
  \doi{10.1006/jcph.1996.0047}.

\bibitem[{Ronneberger et~al.(2015)Ronneberger, Fischer,, and Brox}]{unet}
Ronneberger, O., P.~Fischer, and T.~Brox, 2015: {U-Net}: Convolutional networks
  for biomedical image segmentation. \textit{Medical Image Computing and
  Computer-Assisted Intervention -- MICCAI 2015}, N.~Navab, J.~Hornegger, W.~M.
  Wells, and A.~F. Frangi, Eds., Springer International Publishing, Cham,
  234--241, \doi{10.1007/978-3-319-24574-4_28}.

\bibitem[{Ruffault et~al.(2020)}]{Ruffault2020}
Ruffault, J., and Coauthors, 2020: Increased likelihood of heat-induced large
  wildfires in the {Mediterranean Basin}. \textit{Scientific Reports},
  \textbf{10}, 13\,790, \doi{10.1038/s41598-020-70069-z}.

\bibitem[{Scher and Messori(2021)Scher, and Messori}]{Scher2021}
Scher, S., and G.~Messori, 2021: Ensemble methods for neural network-based
  weather forecasts. \textit{Journal of Advances in Modeling Earth Systems},
  \textbf{13}, \doi{10.1029/2020MS002331}.

\bibitem[{Slingo and Palmer(2011)Slingo, and Palmer}]{Slingo2011}
Slingo, J., and T.~Palmer, 2011: {Uncertainty in weather and climate
  prediction}. \textit{Philosophical Transactions of the Royal Society A:
  Mathematical, Physical and Engineering Sciences}, \textbf{369~(1956)},
  4751--4767, \doi{10.1098/rsta.2011.0161}.

\bibitem[{Sundararajan and Agrawal(2021)Sundararajan, and Agrawal}]{rain_check}
Sundararajan, M., and S.~Agrawal, 2021: The rain check. \textit{4th Workshop on
  Visualization for AI Explainability},
  \urlprefix\url{http://raincheck.karyk.com/rain-check}.

\bibitem[{Sundararajan et~al.(2017)Sundararajan, Taly,, and
  Yan}]{Sundararajan2017}
Sundararajan, M., A.~Taly, and Q.~Yan, 2017: Axiomatic attribution for deep
  networks. \doi{10.48550/arxiv.1703.01365}.

\bibitem[{Sánchez-Benítez et~al.(2018)Sánchez-Benítez, García-Herrera,
  Barriopedro, Sousa,, and Trigo}]{sanchez_benitez_2018}
Sánchez-Benítez, A., R.~García-Herrera, D.~Barriopedro, P.~M. Sousa, and
  R.~M. Trigo, 2018: June 2017: The earliest {European} summer mega-heatwave of
  reanalysis period. \textit{Geophysical Research Letters}, \textbf{45},
  1955--1962, \doi{10.1002/2018GL077253}.

\bibitem[{Sønderby et~al.(2020)}]{sonderby2020metnet}
Sønderby, C.~K., and Coauthors, 2020: {MetNet}: A neural weather model for
  precipitation forecasting. arXiv, \doi{10.48550/ARXIV.2003.12140}.

\bibitem[{Teng et~al.(2013)Teng, Branstator, Wang, Meehl,, and
  Washington}]{Teng2013}
Teng, H., G.~Branstator, H.~Wang, G.~A. Meehl, and W.~M. Washington, 2013:
  Probability of {US} heat waves affected by a subseasonal planetary wave
  pattern. \textit{Nat. Geosci.}, \textbf{6~(12)}, 1056--1061,
  \doi{10.1038/ngeo1988}.

\bibitem[{Ullrich et~al.(2016)Ullrich, Devendran,, and Johansen}]{ulrich_2015b}
Ullrich, P.~A., D.~Devendran, and H.~Johansen, 2016: Arbitrary-order
  conservative and consistent remapping and a theory of linear maps: Part {II}.
  \textit{Monthly Weather Review}, \textbf{144~(4)}, 1529 -- 1549,
  \doi{10.1175/MWR-D-15-0301.1}.

\bibitem[{Ullrich and Taylor(2015)Ullrich, and Taylor}]{ulrich_2015a}
Ullrich, P.~A., and M.~A. Taylor, 2015: Arbitrary-order conservative and
  consistent remapping and a theory of linear maps: Part {I}. \textit{Monthly
  Weather Review}, \textbf{143~(6)}, 2419 -- 2440,
  \doi{10.1175/MWR-D-14-00343.1}.

\bibitem[{Vautard et~al.(2007)}]{Vautard2007}
Vautard, R., and Coauthors, 2007: Summertime {European} heat and drought waves
  induced by wintertime {Mediterranean} rainfall deficit. \textit{Geophysical
  Research Letters}, \textbf{34}, \doi{10.1029/2006GL028001}.

\bibitem[{Vitart et~al.(2017)}]{Vitart2017}
Vitart, F., and Coauthors, 2017: The subseasonal to seasonal ({S2S}) prediction
  project database. \textit{Bulletin of the American Meteorological Society},
  \textbf{98}, 163--173, \doi{10.1175/BAMS-D-16-0017.1}.

\bibitem[{Wang et~al.(2011)Wang, Dolman,, and Alessandri}]{Wang2011}
Wang, G., A.~J. Dolman, and A.~Alessandri, 2011: A summer climate regime over
  {Europe} modulated by the north {Atlantic} oscillation. \textit{Hydrology and
  Earth System Sciences}, \textbf{15}, 57--64, \doi{10.5194/hess-15-57-2011}.

\bibitem[{Weyn et~al.(2020)Weyn, Durran,, and Caruana}]{Weyn2020}
Weyn, J.~A., D.~R. Durran, and R.~Caruana, 2020: Improving data-driven global
  weather prediction using deep convolutional neural networks on a cubed
  sphere. \textit{Journal of Advances in Modeling Earth Systems}, \textbf{12},
  e2020MS002\,109, \doi{10.1029/2020MS002109}.

\bibitem[{Weyn et~al.(2021)Weyn, Durran, Caruana,, and
  Cresswell-Clay}]{Weyn2021}
Weyn, J.~A., D.~R. Durran, R.~Caruana, and N.~Cresswell-Clay, 2021:
  Sub-seasonal forecasting with a large ensemble of deep-learning weather
  prediction models. \textit{Journal of Advances in Modeling Earth Systems},
  \textbf{13~(7)}, e2021MS002\,502, \doi{10.1029/2021MS002502}.

\bibitem[{White et~al.(2017)}]{White2017}
White, C.~J., and Coauthors, 2017: Potential applications of
  subseasonal-to-seasonal ({S2S}) predictions. \textit{Meteorological
  Applications}, \textbf{24}, 315--325, \doi{10.1002/met.1654}.

\bibitem[{White et~al.(2021)White, Kornhuber, Martius,, and Wirth}]{white2021}
White, R.~H., K.~Kornhuber, O.~Martius, and V.~Wirth, 2021: From atmospheric
  waves to heatwaves: A waveguide perspective for understanding and predicting
  concurrent, persistent and extreme extratropical weather. \textit{Bulletin of
  the American Meteorological Society}, 1--35, \doi{10.1175/BAMS-D-21-0170.1}.

\bibitem[{Wilks(2019)}]{Wilks2019}
Wilks, D.~S., 2019: \textit{Statistical Methods in the Atmospheric Sciences}.
  4th ed., Elsevier, 369-483 pp., \doi{10.1016/B978-0-12-815823-4.00009-2}.

\bibitem[{Wright et~al.(2014)Wright, de~Beurs,, and Henebry}]{Wright2014}
Wright, C.~K., K.~M. de~Beurs, and G.~M. Henebry, 2014: Land surface anomalies
  preceding the 2010 {Russian} heat wave and a link to the north {Atlantic}
  oscillation. \textit{Environmental Research Letters}, \textbf{9}, 124\,015,
  \doi{10.1088/1748-9326/9/12/124015}.

\bibitem[{Wu et~al.(2002)Wu, Geller,, and Dickinson}]{Wu2002}
Wu, W., M.~A. Geller, and R.~E. Dickinson, 2002: The response of soil moisture
  to long-term variability of precipitation. \textit{Journal of
  Hydrometeorology}, \textbf{3~(5)}, 604--613,
  \doi{10.1175/1525-7541(2002)003<0604:TROSMT>2.0.CO;2}.

\bibitem[{Wulff and Domeisen(2019)Wulff, and Domeisen}]{Wulff2019}
Wulff, C.~O., and D.~I.~V. Domeisen, 2019: Higher subseasonal predictability of
  extreme hot {European} summer temperatures as compared to average summers.
  \textit{Geophysical Research Letters}, \textbf{46}, 11\,520--11\,529,
  \doi{10.1029/2019GL084314}.

\bibitem[{Xu et~al.(2016)Xu, FitzGerald, Guo, Jalaludin,, and Tong}]{Xu2016}
Xu, Z., G.~FitzGerald, Y.~Guo, B.~Jalaludin, and S.~Tong, 2016: Impact of
  heatwave on mortality under different heatwave definitions: A systematic
  review and meta-analysis. \textit{Environment International}, \textbf{89-90},
  193--203, \doi{10.1016/j.envint.2016.02.007}.

\bibitem[{Yosinski et~al.(2014)Yosinski, Clune, Bengio,, and
  Lipson}]{yosinski_2014}
Yosinski, J., J.~Clune, Y.~Bengio, and H.~Lipson, 2014: How transferable are
  features in deep neural networks? \textit{Advances in Neural Information
  Processing Systems}, Z.~Ghahramani, M.~Welling, C.~Cortes, N.~Lawrence, and
  K.~Weinberger, Eds., Curran Associates, Inc., Vol.~27,
  \urlprefix\url{https://proceedings.neurips.cc/paper/2014/hash/375c71349b295fbe2dcdca9206f20a06-Abstract.html}.

\bibitem[{Yu and Koltun(2016)Yu, and Koltun}]{dilated_conv}
Yu, F., and V.~Koltun, 2016: Multi-scale context aggregation by dilated
  convolutions. \textit{4th International Conference on Learning
  Representations, {ICLR} 2016, San Juan, Puerto Rico, May 2-4, 2016,
  Conference Track Proceedings}, Y.~Bengio, and Y.~LeCun, Eds.,
  \doi{10.48550/arXiv.1511.07122}.

\bibitem[{Zhou et~al.(2019)Zhou, Yang, Chen, Zhang, Huang,, and La}]{Zhou2019}
Zhou, Y., B.~Yang, H.~Chen, Y.~Zhang, A.~Huang, and M.~La, 2019: Effects of the
  madden–julian oscillation on 2-m air temperature prediction over china
  during boreal winter in the s2s database. \textit{Climate Dynamics},
  \textbf{52}, 6671--6689, \doi{10.1007/s00382-018-4538-z}.

\bibitem[{Zhu et~al.(2002)Zhu, Toth, Wobus, Richardson,, and Mylne}]{Zhu2002}
Zhu, Y., Z.~Toth, R.~Wobus, D.~Richardson, and K.~Mylne, 2002: The economic
  value of ensemble-based weather forecasts. \textit{Bulletin of the American
  Meteorological Society}, \textbf{83}, 73--84,
  \doi{10.1175/1520-0477(2002)083<0073:TEVOEB>2.3.CO;2}.

\end{thebibliography}

\end{document}


\maketitle


%
%
%

%



\section{Interhemispheric differences in model skill}

\begin{figure}[h]
 \noindent\includegraphics[width=39pc]{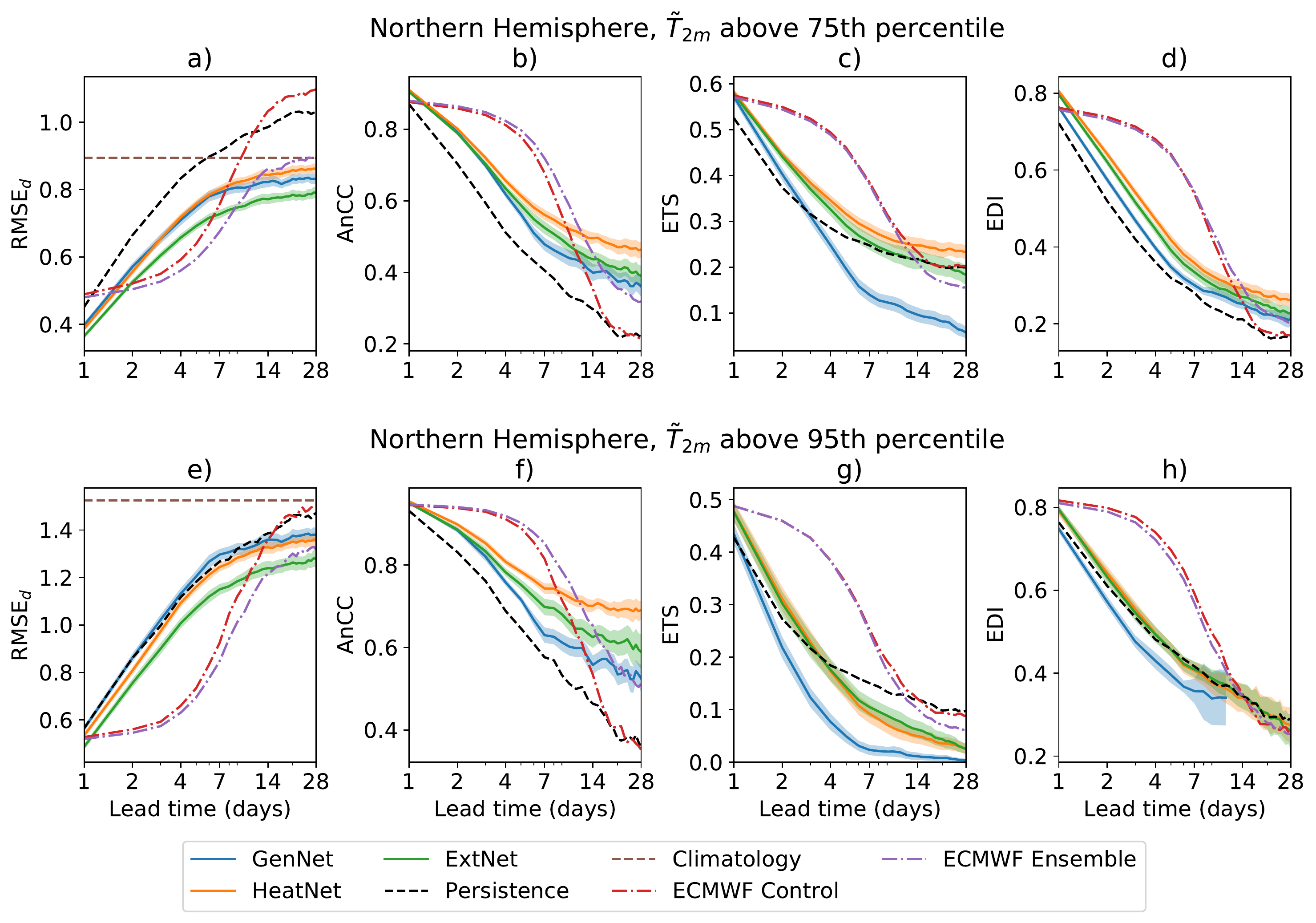}\\
 \caption{Forecast metrics for different models during the summer months of 2017-2021 and over land in the Northern Hemisphere. Metrics are shown for forecasts conditioned on targets being above the 75th (top) and 95th (bottom) percentiles of standardized temperature anomalies over land during summers in the test set. From left to right, the debiased root mean squared error (RMSE$_d$), centered anomaly correlation coefficient (AnCC), equitable threat score (ETS) and extremal dependence index (EDI) are shown. Uncertainty bands, shown for the NWMs as a reference, represent 1 standard deviation. Results are only shown for metrics with a robust uncertainty estimate; details may be found in Appendix A.
 }
 \label{fig:nh_metrics_with_lead_time}
\end{figure}

\begin{figure}[h]
 \noindent\includegraphics[width=39pc]{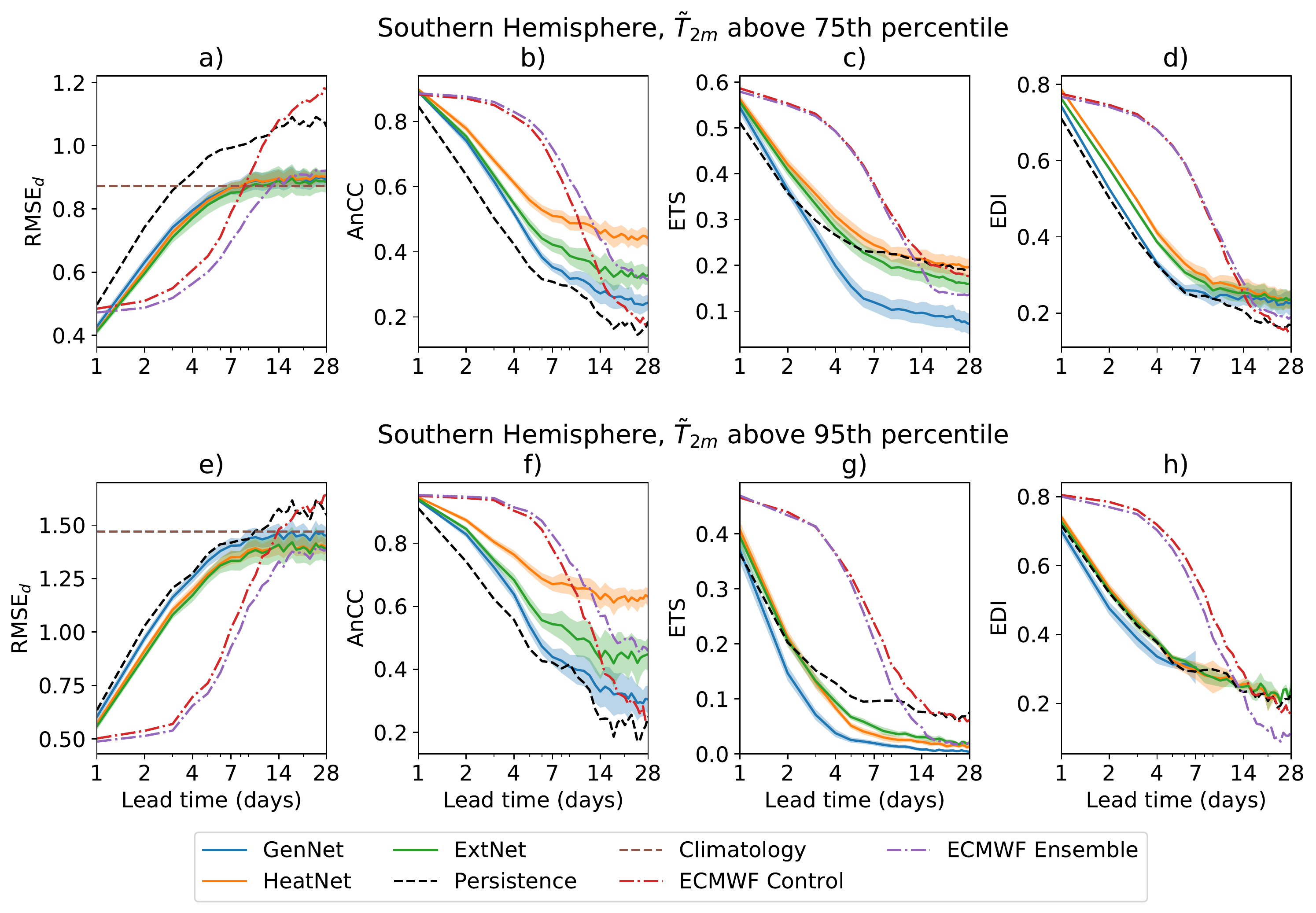}\\
 \caption{Same as Figure \ref{fig:nh_metrics_with_lead_time} but for land in the Southern Hemisphere.}
 \label{fig:sh_metrics_with_lead_time}
\end{figure}

\blfootnote{Corresponding author: Ignacio Lopez-Gomez, ilopezgp@google.com}
Figures \ref{fig:nh_metrics_with_lead_time} and \ref{fig:sh_metrics_with_lead_time} show the same metrics as Figure 3 in the main text, but computed in this case over the Northern and Southern Hemispheres, respectively. All neural weather models perform better in the Northern Hemisphere than in the Southern Hemisphere. This skill difference is less pronounced in the physics-based models; for instance, ExtNet retains higher correlations with the targets over the Northern Hemisphere than the ECMWF ensemble mean after two weeks, but over the Southern Hemisphere the physics-based ensemble remains more accurate at all lead times.

\clearpage

%






%


